\documentclass[11pt,reqno]{amsart}
\usepackage{graphicx}
\usepackage{subfigure}
\usepackage{amsmath}
\usepackage{amsfonts}
\usepackage{amssymb}
\usepackage{scrextend}
\usepackage[labelsep=space]{caption}
\usepackage[font=footnotesize]{caption}
\usepackage{caption}
\usepackage{cite}
\usepackage{xcolor}
\usepackage{hyperref}
\usepackage{multirow}
\usepackage{morefloats}
\usepackage{array}
\newcommand{\bd}{\begin{dfn}}
	\newcommand{\ed}{\end{dfn}}
\newcommand{\br}{\begin{rmk}}
	\newcommand{\er}{\end{rmk}}
\newcommand{\bt}{\begin{thm}}
	\newcommand{\et}{\end{thm}}
\newcommand{\bcl}{\begin{clry}}
	\newcommand{\ecl}{\end{clry}}
\newcommand{\bl}{\begin{lem}}
	\newcommand{\el}{\end{lem}}
\newcommand{\bp}{\begin{proof}}
	\newcommand{\ep}{\end{proof}}

\nonstopmode \numberwithin{equation}{section}
\newtheorem{thm}{Theorem}[section]
\newtheorem{lem}[thm]{Lemma}
\newtheorem{theorem}{Theorem}[section]
\newtheorem{clry}[theorem]{Corollary}
\newtheorem{dfn}[thm]{Definition}
\newtheorem{rmk}[thm]{Remark}

\title[]{Approximate Analytical Solution using Power Series Method for the Propagation of Blast Waves in a Rotational Axisymmetric non-ideal Gas}
\author[]{}
\begin{document}
	\leftline{ \scriptsize \it  }
	\keywords{Shock and blast waves, Rankine-Hugoniot conditions, Power series solutions, Non-ideal gas, Rotating medium}
	
	\maketitle
	\begin{center}
		\textbf{Nandita} \footnote{\textbf{ Email:} nandita@as.iitr.ac.in, nanditagupta1214@gmail.com }, \textbf{Rajan Arora}\footnote{ \textbf{Email:} rajan.arora@as.iitr.ac.in}\\
		\textsuperscript{1,2} Department of Applied Mathematics and Scientific Computing\\
		Indian Institute of Technology Roorkee, India\\
	\end{center}
	\begin{abstract}
		In this paper, the propagation of the blast (shock) waves in non-ideal gas atmosphere in rotational medium is studied using a power series method in cylindrical geometry. The flow variables are assumed to be varying according to the power law in the undisturbed medium with distance from the symmetry axis. To obtain the similarity solution, the initial density is considered as constant in the undisturbed medium. Approximate analytical solutions are obtained using Sakurai's method by extending the power series of the flow variables in power of ${\left( {\frac{{{a_0}}}{U}} \right)^2}$, where $U$ and $a_0$ are the speeds of the shock and sound, respectively, in undisturbed fluid. The strong shock wave is considered for the ratio ${\left( {\frac{{{a_0}}}{U}} \right)^2}$ which is considered to be a small quantity. With the aid of that method, the closed-form solutions for the zeroth-order approximation is given as well as first-order approximate solutions are discussed. Also, with the help of graphs behind the blast wave for the zeroth-order approximation, the distributions of variables such as density, radial velocity, pressure and azimuthal fluid velocity are analyzed. The results for the rotationally axisymmetric non-ideal gas environment are compared to those for the ideal gas atmosphere.
	\end{abstract}
	\section{\textbf{Introduction}}
	The phenomenon of shock waves has received a lot of attention among researchers of various fields. Explosions, projectiles at supersonic speeds or aircrafts traveling at supersonic speed or moving bullets are some examples of shock waves. Whether it's an explosion, a high-speed bullet, or a high-speed aeroplane, these kinds of phenomena create massive changes in pressure, temperature, and density in a very short period of time. Due to research progress in this field, the study of shock waves is now very useful in geophysics, plasma physics, nuclear science, astrophysics and interstellar gas masses. Shock waves are common in the interstellar medium due to a wide variety of high-speed movements and intense events, such as cloud collisions, bipolar outbursts from young protostars, massive mass losses by huge stars in the later stages of their emergence, supernova explosions, the mid-region of starburst galaxies, and more. A disturbance that's created by a strong shock wave called a ``blast wave" is produced in the surrounding medium when a significant amount of energy is rapidly released into a small space, such as during an explosion or spark discharge in air. A forward-moving shock wave arises once the ``blast wave" process occurs (see \cite{1,2,3}). ``Blast waves" will have various geometries depending on the spatial distribution of the supernovae.
	
	After the explosion, the shock front, a surface around the blast wave that flows with diminishing speed in an outward direction, releases a higher force. As a result, the medium does not follow the ideal condition. Therefrore, in such a situation, a non-ideal medium and its effects should be considered. The blast wave solutions have been studied for a variety of parameters by Guderley \cite{4}, where the higher-order terms of sound and shock front speed quotient ${\left( {\frac{{{a_0}}}{U}} \right)^2}$ being neglected as $\left(a_0<<U\right)$ for the strong shock. Taylor \cite{5,6} has investigated the blast waves produced by the massive explosion. In order to find first and second-order approximations of solutions for planar and cylindrical shock waves, Sakurai then expanded on Taylor's work \cite{5} by taking into account the power series technique (see \cite{7,8}). Siddiqui et al. \cite{9} and Nath and Singh \cite{10} investigated the propagation of blast waves in magnetogasdynamics under both ideal and non-ideal circumstances using Sakurai's technique. An explicit solution for the gas dynamics equations was observed by McVittie \cite{11} considering the three-dimensional spherical geometry. The formation of self-similar problems and instances describing the adiabatic motion of a gas model of stars in a non-rotating medium were examined in \cite{12,13}. Also, many literature \cite{14,15,16,17,18,35} obtained the approximate analytical solution for a blast wave propagation in a rotating medium using power series method. Using Lie group technique, the analytical solution of imploding shocks in a Non-Ideal Gas was obtained by Chauhan \cite{19} and Singh et al. \cite{34}. Macroscopic motion occurs in the interplanetary environment at supersonic speeds, resulting in shock waves. Considering the 1D problem, the point-like strong explosions problem has been thoroughly studied in \cite{5,22}. Sedov's explosion problem \cite{22} is still used for simple modelling of supernova remnants in their initial evolutionary states (see \cite{23,24,25,26,27}). Non-linear models are used to model a range of physical processes that occur around us. Determining analytical solutions  to non-linear equations is a difficult task, and we discover the solutions in closed form only in a few cases \cite{28,29}. With the help of Lie group transformation, for a strong converging shock, the similarity solutions in a non-ideal gas were found by Arora et al.\cite{30}, the effects of non-ideal magnetogasdynamics under monochromatic radiation were shown by Sharma and Arora \cite{31}, for the isothermal flow the strong cylindrical shock wave in a self-gravitating non-ideal gas with the influence of axial magnetic field have been investigated by Gupta et al. \cite{32}, self-similar solutions to the problem of shock wave propagation through reacting polytropic gases by considering burnt and unburnt gas have been obtained by Jena and Singh \cite{33}. The evolution equation for spherical shock wave in an interstellar van der Waals gas cloud is derived by Singh et al.\cite{37}. An analytical solution for a cylindrical blast wave that is expanding with transition from the strong to the weak shock has been studied by Raga et al. \cite{36}. As a result, the literature has a significant number of research articles on the propagation of shock waves in gas dynamics.
	
	In terms of solar activity, shock waves produced by solar flares are the most powerful form that can be observed in the field of solar-terrestrial physics. Shock waves distribute more than half of the energy released in small or large flares. Any flare ejecta-driven shock would be characterised as a piston-driven shock with a radially dependent shock velocity, unlike blast-produced shock $U={A^{*}}{r_{s}}^{\delta}$ (see previous studies \cite{40,41}). The creation of parallel methods for solving the astrophysical phenomena with large spatial and time scales is a prerequisite for more complex calculations. Rybakin et al. \cite{42} devised a parallel programme to simulate the fragmentation and creation of filamentous structures in molecular clouds, in which new computational codes have been used  when shock waves of different types interact with unevenly distributed molecular clouds in galaxies. Shock waves' propagation, collision, and interaction with molecular clouds all contribute to the sequence of events that results in star formation through self-gravity of gas in dynamically fluctuating molecular clouds. One of the fundamental issues in astrophysics is the explanation and investigation of the internal motion in stars. Observational data show that flare-ups in novae and supernovae are caused by the sudden emission of energy and the unsteady motion of a significant amount of gas. The equations of motion and equilibrium can be used to analyse the qualitative behaviour of the gaseous mass. The results of a numerical simulation of a supernova strong shock interaction with an interstellar molecular cloud in three dimensions have been published by Rybakin et al. \cite{43}. They modelled the development of hydrodynamical flow, contraction, fragmentation processes, and turbulent flow generation in the cloud and surrounding media without taking into account the effect of rotation. Rybakin et al. \cite{44} has also studied the processes of molecular cloud collisions in three-dimensional numerical simulations.
	
	The majority of publications on shock or blast waves have not found an analytical solution for blast wave propagation in a rotating medium. The flow behind blast (shock) wave caused by a strong explosion in a rotating non-ideal gas has self-similar analytical solutions, which we have obtained in the current study. By considering the impact of azimuthal fluid velocity and rotational parameter, we generalised the work of Sakurai \cite{7}. In the present study, we are considering that, immediately following the explosion, the surface of shock wave surrounds the front of the blast wave. The distribution of hydrodynamical quantities behind the shock front constantly varies as it moves with decreasing velocity. The azimuthal component of fluid velocity are intended to change and follow power laws, but the initial density is assumed to be constant. When the shock speed is high enough to ignore the ratio ${\left({\frac{{{a_0}}}{U}} \right)^2}$, our result corresponds to the approximation validity, where speed of sound and shock front are represented by $a_0$ and $U$ respectively. As a result, the solution in the form of a power series in ${\left( {\frac{{{a_0}}}{U}} \right)^2}$ is presented in this work. Tables and figures are used to discuss the impact of the adiabatic exponent, non-ideal gas parameter and rotational parameter on flow variables and shock waves. The values of the non-ideal gas parameter $b$ are taken from the published paper by Singh et al.\cite{39} and the values of the adiabatic exponent $\gamma$ and non-rotating parameter $\frac{{{v^*}}}{{{A^*}}}$ are taken from Nath et al. \cite{50}. A comparison of the solution is also presented between the non-rotating and rotating medium for both ideal and non-ideal gas.
	
	\section{\textbf{Equations of Motion and Boundary Conditions}}
	We study the blast wave propagation for one-dimensional rotational axisymmetric unsteady flow of non-ideal gas by considering the azimuthal components of fluid velocity. ``The fundamental equations are given as (see \cite{15,16,17,18,19,22,23,50}).
	\begin{align}\label{1}
		&{\rho _t} + u{\rho _r} + \rho {u_r} + \frac{{\rho u}}{r} = 0,
	\end{align}
	\begin{align}\label{2}
		&{u_t} + u{u_r} + \frac{1}{\rho }{p_r} - \frac{{{v^2}}}{r} = 0,
	\end{align}
	\begin{align}\label{3}
		&{p_t} + u{p_r} + \rho {a^2}\left( {{u_r} + \frac{u}{r}} \right) = 0,
	\end{align}
	\begin{align}\label{4}
		&{v_t} + u{v_r} + \frac{{uv}}{r} = 0,
	\end{align} 
	\\		
	where $\rho$ denotes the density, $p$ the pressure, $v$ and $u$ the azimuthal and radial components of the fluid velocity $\vec q$ in coordinates $(r,\theta,z)$, respectively. The independent variables are the space coordinate $r$ and time $t$. The non-numeric subscripts denote the partial differentiation with respect to the indicated variables unless stated otherwise. 
	\\	
	Now, by assuming that the gas obeys a simplified van der Waals equation of state of form (see \cite{32,45,46} ), the equation of state and the internal energy of non-ideal gas per unit mass is given as
	\begin{equation}\label{5}
		p = \frac{R\rho T}{ {1 - b\rho } },\quad 
		{\rm E_m}=c_vT = \frac{{p\left( {1 - b\rho } \right)}}{{\rho \left( {\gamma  - 1} \right)}}, 
	\end{equation}
	where internal energy per unit mass is denoted by $\rm E_m$, $\gamma$ is the ratio of the specific heats of the gas. For the most of the van der waals gases, the value of the specific heat ratio of the gas lies between $\left( {1 < \gamma  < 2} \right)$, $b$ is the van der Waals excluded volume which places the limit, $\rho_{max}=\frac{1}{b}$, on the density of the gas, but at high temperature it tends to be a constant value equal to the internal volume of the gas molecules which lies in the range $0.9 \times {10^{ - 3}} \le b \le 1.1 \times {10^{ - 3}}$, $R$ is the universal gas constant and $T$ is the absolute temperature, $c_v$ is specific heat at constant volume.
	Also, the velocity of sound `$a$'of the non-ideal gas is given by
	\begin{equation}\label{6}	
		a = \left\{\frac{{\gamma p}}{{\rho \left( {1 - b\rho } \right)}}\right\}^{\frac{1}{2}}.
	\end{equation}
	\\
	The azimuthal component of fluid velocity $v$ and the angular velocity of the medium `$A$' at radial distance `$r$' from the axis of symmetry are related by the relation
	\begin{equation}\label{7}	
		v=Ar.
	\end{equation}
	The system of equations \eqref{1}-\eqref{4} can be rewritten in conservative form as follows:
	\begin{equation}\label{8}
		{G_t}\left( {r,t,\textbf{E}} \right) + {H_r}\left( {r,t,\textbf{E}} \right) = I\left( {r,t,\textbf{E}} \right).
	\end{equation}
	Here, $\textbf{E} = {\left( {\rho ,u,p,v} \right)^T}$ is the column vector of the primitive variables, and $G$, $H$ and $I$ are as follows:	
	\begin{equation}\label{9}
		\left.
		\begin{aligned}
			&G = {\left( {\rho ,\rho u,\frac{{\rho {u^2}}}{2} + \frac{{p\left( {1 - b\rho } \right)}}{{\gamma  - 1}},\rho v} \right)^T},\\
			&H = {\left( {\rho u,\rho {u^2} + p,\left( {\frac{{\rho {u^2}}}{2} + \frac{{p\left( {1 - b\rho } \right)}}{{\gamma  - 1}} + pu} \right)u,\rho vu} \right)^T},\\
			&I = {\left( { - \frac{{\rho u}}{r},\frac{{\rho {v^2}}}{r} - \frac{{\rho {u^2}}}{r},\frac{{\rho u{v^2}}}{r} - \frac{{\rho {u^3}}}{{2r}} - \frac{{p\left( {\gamma  - b\rho } \right)u}}{{\left( {\gamma  - 1} \right)r}}, - \frac{{2\rho uv}}{r}}\right)^T}.
		\end{aligned}
		\right\}
	\end{equation}
	Then, the Rankine-Hugoniot condition for shock waves is given by
	\begin{equation}\label{10}
		U\left[ {{G_j}} \right] = \left[ {{H_j}} \right], 
	\end{equation}
	where $j=1,2,3,4$ and $U$ denotes velocity of the shock front. Also, $[F(x)]=F(x_1)-F(x_0)$ is showing the jump for any quantity $F(x)$, and the medium just ahead and behind the shock front is denoted by $x_0$ and $x_1$, respectively.\\
	
	It is assumed that a diverging blast wave is propagating outwards from the axis of symmetry with velocity $U=\frac{dr_s}{dt}$ in the rotating medium with constant density, zero radial velocity, and variable azimuthal velocity. The flow variables just ahead of the shock front are as follows:
	\begin{equation}\label{11}
		u=u_0=0,\quad
		\rho=\rho_{0},\quad
		v=v_0={v^{*}}{r_{s}}^{\alpha},
	\end{equation}
	where $r_s$ is the radius of the shock, $v^{*}$ and $\alpha$ are constants, the conditions just ahead of the shock front are represented by the subscript `0'. The relation between radius of the shock $r_s$ and the velocity of the shock front $U$ is given by \cite{17,18,20}
	\begin{equation}\label{12}
		U={A^{*}}{r_{s}}^{\delta},
	\end{equation}
	where $A^{*}$ and $\delta$ are constant.
	
	From Eqs. $\eqref{7}$ and $\eqref{11}$, we obtained the initial angular velocity of the medium as:
	\begin{equation}\label{13}
		A_{0}=v^{*}{r_{s}}^{\alpha-1},
	\end{equation}
	Also, from Eqs. \eqref{2},\eqref{7} and \eqref{13} we find 
	\begin{equation}\label{14}
		p_{0}=\frac{{v^{*}}^{2}{r_{s}}^{2{\alpha}}{\rho_{0}}}{2{\alpha}},\quad{\alpha\ne 0}.
	\end{equation}
	The constants $\alpha$ and $\delta$, which are exponents in Eq. \eqref{11} and \eqref{12}, respectively, necessarily satisfy the relation $\alpha=\delta$ (see the required condition in Eq. \eqref{28}) for the similarity solution to exist. These constants can take arbitrary values as $\alpha>0$ (see Eq. \eqref{14}). In addition, the initial angular velocity given in equation \eqref{13} influences the choice of $\alpha$. If $\alpha=1$, then the initial angular velocity remains constant. If the distance from the axis increases, the initial angular velocity drops. As a result, the physical significance of these constants is found in the range $0<\alpha=\delta \le 1$.
	\\
	Now, from the conservative form $\eqref{8}$ and the Rankine-Hugoniot jump condition $\eqref{10}$, the jump conditions for the strong shock across the shock front $r=r_{s}(t)$ are given as \cite{7,8}: 	
	\begin{equation}\label{15}
		\left.
		\begin{aligned}
			&{\rho _1}(r_{s}(t),t) = \frac{{{\rho _0}\left( {\gamma  + 1} \right)}}{{\left( {\gamma  - 1} \right)}}\left\{ 1 - \frac{{2{b}{\rho_{0}}}}{{\gamma  - 1}} - \frac{2{\left( {1 - b{\rho_{0}}} \right)}}{\gamma  - 1}{\frac{a_0^2}{U^2}}\right\},\\
			&{u_1}(r_{s}(t),t) = \frac{2U}{\gamma  + 1}\left\{{1 - b{\rho_{0}} - {\left( {1 - b{\rho_{0}}} \right)}{\frac{a_0^2}{U^2}}} \right\},\\
			&{p_1}(r_{s}(t),t) = \frac{{2{\rho _0}{U^2}}}{\gamma  + 1}\left\{ {1 - b{\rho_{0} - \frac{{\left( {1 - b{\rho_{0}}} \right)\left( {\gamma  - 1} \right)}}{2\gamma}\frac{{a_0^2}}{{{U^2}}}}} \right\},\\
			&{v_1}(r_{s}(t),t) = {v_0}.
		\end{aligned}
		\right\}
	\end{equation}
	\\
	where $a_{0}^{2}=\frac{{\gamma {p_0}}}{{{\rho_{0}} \left( {1 - b{\rho_{0} }} \right)}}$ is the speed of sound in the undisturbed medium. 
	The density ratio $\psi=\frac{\rho_0}{\rho_1}$ is obtained as:
	\begin{equation}\label{16}
		\frac{{{\rho _0}}}{{{\rho _1}}} = \frac{{\left( {\gamma  - 1} \right)}}{{\left( {\gamma  + 1} \right)}} + \frac{2b{\rho_{0}}}{{\left( {\gamma  + 1} \right)}} + \frac{{2\left( {1 - b{\rho_{0}}} \right)}}{{\left( {\gamma  + 1} \right)}}\frac{{a_0^2}}{{{U^2}}}.
	\end{equation} 
	
	Also, the energy generated by the blast wave is equal to the energy released by the explosion and thus remains constant \cite{7}. Consequently, the energy balance equation is as follows::
	\begin{equation}\label{17}
		E = \int_0^{{r_s}} {\left[ {\frac{1}{2}\left( {{u^2} + {v^2}} \right) - \frac{1}{2}\left( {v_0^2} \right) + \frac{1}{{\left( {\gamma  - 1} \right)}}\left\{ {\frac{{\left( {1 - b\rho } \right)p}}{\rho } - \frac{{\left( {1 - b{\rho _0}} \right){p_0}}}{{{\rho _0}}}} \right\}} \right]\rho rdr,}
	\end{equation} 
	where $E$ is the explosion energy per unit area of the shock front of blast wave when $r_{s}=1$." By considering the Lagrangian continuity equation, we have
	\begin{equation}\label{18}
		\int_0^{{r_s}} {\frac{\rho }{{{\rho _0}}}{r}dr = \frac{{{r_s}^{2}}}{{2}},}
	\end{equation}
	From equations \eqref{13},\eqref{14} and \eqref{19}, we transformed the expression given in \eqref{18} as follows: 
	\begin{equation}\label{19}
		E = \int_0^{r_s} {\left\{ {\frac{\rho }{2}\left( {{u^2} + {v^2}} \right) + \frac{\left( {1 - b\rho } \right)p}{\left( {\gamma  - 1} \right)}} \right\}{r}dr - \left\{{\frac{\rho _0}{2}\left( {v_0^2} \right) + \frac{\left( {1 - b{\rho_0}} \right){p_0}}{\left( {\gamma  - 1} \right)}} \right\}\frac{r_s^{2}}{2}}. 
	\end{equation}
	\section{\textbf{Transformation of the Basic Equations}} 	 
	Now, we are introducing $x$ and $y$ in place of $r$ and $t$ as new dimensionless independent variables defined by (see \cite{7})
	\begin{equation}\label{20}
		\frac{r}{r_s} = x,\quad
		{\left( {\frac{{{a_0}}}{U}} \right)^2} = y.
	\end{equation}
	Using \eqref{20}, we express all the physical quantities $\rho$, $u$, $p$ and $v$ as follows:
	\begin{equation}\label{21}
		\left.
		\begin{aligned}
			&{\rho _1} = {\rho _0}\pi \left( {x,y} \right),\\
			&{u_1} = Uf\left( {x,y} \right),\\
			&{p_1} = {p_0}{\left( {\frac{U}{{{a_0}}}} \right)^2}g\left( {x,y} \right)=\frac{{p_0}{g\left({x,y}\right)}}{y},\\
			&{v_1} = U\phi \left( {x,y} \right),
		\end{aligned}
		\right\}
	\end{equation}
	where $\pi$, $f$, $g$ and $\phi$ are non-dimensional quantities and functions of $x$ and $y$. The operators using \eqref{20} and \eqref{21} are given as:  
	\begin{equation}\label{22}
		\frac{\partial}{{\partial r}} = \frac{1}{{{r_s}}}\frac{\partial }{{\partial x}},\quad
		\frac{D}{{Dt}} =\frac{\partial}{{\partial t}}+u\frac{\partial}{{\partial r}}= \frac{U}{{{r_s}}}\left\{ {\left( {f - x} \right)\frac{\partial }{{\partial x}} + \lambda y\frac{\partial }{{\partial y}}} \right\},\\
	\end{equation}
	where $\lambda  = {{{r_s}\left( {\frac{{dy}}{{d{r_s}}}} \right)} \mathord{\left/{\vphantom {{{r_s}\left( {\frac{{dy}}{{d{r_s}}}} \right)} y}} \right.\kern-\nulldelimiterspace} y}{\rm{ }}$ is function of $y$ only. The following system of equations is obtained by applying the operators \eqref{22} in the fundamental equations \eqref{1}-\eqref{4} along with the flow variable values from \eqref{21}:
	\begin{align}
		&\left( {f - x} \right){\pi _x} + \lambda y{\pi _y} + \pi \left( {{f_x} + \frac{f}{x}} \right) = 0,\label{23}\\
		&\left( {f - x} \right)\pi {f_x} + \lambda y\pi {f_y} + \frac{{1 - b{\rho _0}}}{\gamma }{g_x} = \frac{{\lambda \pi f}}{2} + \frac{{\pi {\phi ^2}}}{x},\label{24}\\
		&\left( {f - x} \right){g_x} + \lambda y{g_y} + \frac{{\gamma g}}{{\left( {1 - b{\rho _0}\pi } \right)}}\left( {{f_x} + \frac{f}{x}} \right) = \lambda g,\label{25}\\
		&\left( {f - x} \right){\phi _x} + \lambda y{\phi _y} = \frac{{\lambda \phi }}{2} - \frac{{f\phi }}{x}.\label{26}
	\end{align}
	Using \eqref{20} and \eqref{21}, in energy equation \eqref{19}, we obtained
	\begin{align}\label{27}
		y{\left( {\frac{{{r_{{s_0}}}}}{{{r_s}}}} \right)^{2}} = &\int_0^1 {\left\{ {\frac{{\gamma \pi }}{{2{{\left( {1 - b{\rho _0}} \right)}}}}\left( {{f^2} + {\phi ^2}} \right) + \frac{{\left( {1 - b{\rho _0}\pi } \right)g}}{{\left( {\gamma  - 1} \right)}}} \right\}} {x}dx\nonumber
		\\
		&- \left\{ {\frac{y{\left( {1 - b{\rho _0}} \right)}}{{ {2}\left( {\gamma  - 1} \right)}} + \frac{\gamma }{{4{{\left( {1 - b{\rho _0}} \right)}}}}{{{\left( {\frac{{{v^*}}}{{{A^*}}}} \right)}^2}} } \right\},
	\end{align}
	where $r_{s_{0}}={\left(\frac{E}{p_{0}}\right)}^{\frac{1}{2}}$. Using equations \eqref{20} and \eqref{21} in \eqref{15} and by taking $\alpha=\delta$ to get the similarity solution, from the R-H jump conditions \eqref{15} we obtained 
	\begin{equation}\label{28}
		\left.
		\begin{aligned}
			&f\left( {1,y} \right) = \frac{{2\left( {1 -b{\rho _0}} \right)}}{{\gamma  + 1}}\left( {1 - y} \right),\\
			&\pi \left( {1,y} \right) = \frac{{\gamma  + 1}}{{\gamma  - 1}}\left[ {1 - \frac{{2 b{\rho _0}}}{{\gamma  - 1}} - \frac{{2\left( {1 - b{\rho _0}} \right)}}{{\gamma  - 1}}y} \right],\\
			&g\left( {1,y} \right) = \frac{{2\gamma }}{{\gamma  + 1}}\left[ {1 - \frac{{\left( {\gamma  - 1} \right)}}{{2\gamma }}y} \right],\\
			&\phi \left( {1,y} \right) = \frac{{{v^*}}}{{{A^*}}}.
		\end{aligned}
		\right\}
	\end{equation}
	
	Now, differentiating the equation \eqref{27} with respect to $y$ we get the value of $\lambda$ as:
	\begin{equation}\label{29}
		\lambda  = \frac{{2J - \frac{y{\left( {1 - b{\rho _0}} \right)}}{{\gamma  - 1}} - \frac{\gamma }{{2{{\left( {1 - b{\rho _0}} \right)}}}} {{{\left( {\frac{{{v^*}}}{{{A^*}}}} \right)}^2}} }}{{J - y\frac{{dJ}}{{dy}} - \frac{\gamma }{{4{{\left( {1 - b{\rho _0}} \right)}}}} {{{\left( {\frac{{{v^*}}}{{{A^*}}}} \right)}^2}} }},		 
	\end{equation}
	where $J$ is defined as:
	\begin{equation}\label{30}
		J = \int_0^1 {\left[ {\frac{{\gamma \pi }}{{2{{\left( {1 - b{\rho _0}} \right)}}}}\left( {{f^2} + {\phi ^2}} \right) + \frac{{\left( {1 - b{\rho _0}\pi } \right)g}}{{\left( {\gamma  - 1} \right)}}} \right]} {x}dx.\\
	\end{equation}
	\section{\textbf{Formation of Power Series Solutions in y}}
	Since the shock front's velocity $U$ is much greater than the speed of sound $a_ 0$ for strong shock waves, thus the quantity $y$ is assumed to be quite small. ``The non-dimensional functions $\pi$, $f$, $g$, and $\phi$ can therefore be expanded as a convergent series in powers of $y$ as shown below:
	\begin{equation}\label{31}
		\left.
		\begin{aligned}
			&f = {f^{\left( 0 \right)}} + y{f^{\left( 1 \right)}} + {y^2}{f^{\left( 2 \right)}} + ......,\\
			&\pi  = {\pi ^{\left( 0 \right)}} + y{\pi ^{\left( 1 \right)}} + {y^2}{\pi ^{\left( 2 \right)}} + ......,\\
			&g = {g^{\left( 0 \right)}} + y{g^{\left( 1 \right)}} + {y^2}{g^{\left( 2 \right)}} + ......,\\
			&\phi  = {\phi ^{\left( 0 \right)}} + y{\phi ^{\left( 1 \right)}} + {y^2}{\phi ^{\left( 2 \right)}} + ......,
		\end{aligned}
		\right\}
	\end{equation}
	where $\pi^{\left( k \right)}$, $f^{\left( k \right)}$, $g^{\left( k \right)}$ and $\phi^{\left( k \right)}$ are the functions of $x$ only and $k=0,1,2...$. Now, using Eq. \eqref{31} in Eq. \eqref{30}, we obtain
	
	\begin{equation}\label{32}
		J = {J_0}\left( {1 + {\vartheta _1}y + {\vartheta _2}{y^2} + {\vartheta _3}{y^3} + ......} \right)\\
	\end{equation}
	where
	\begin{equation}\label{33}
		J_0 = \int_0^1 {\left[ {\frac{{\gamma {\pi ^{\left( 0 \right)}}}}{{2{{\left( {1 - b{\rho _0}} \right)}}}}\left( {{{\left( {{f^{\left( 0 \right)}}} \right)}^2} + {{\left( {{\phi ^{\left( 0 \right)}}} \right)}^2}} \right) + \frac{{\left( {1 - b{\rho _0}{\pi^{\left(0\right)}} } \right)g^{\left(0\right)}}}{{\left( {\gamma  - 1} \right)}}} \right]} {x}dx,
	\end{equation}
	\begin{align}\label{34}
		{\vartheta _1}{J_0} = \int_0^1
		&\left[ {\frac{\gamma }{{2{{\left( {1 - b{\rho _0}} \right)}}}}\left\{ {2{\pi ^{\left( 0 \right)}}\left( {{f^{\left( 0 \right)}}{f^{\left( 1 \right)}} + {\phi ^{\left( 0 \right)}}{\phi ^{\left( 1 \right)}}} \right) + {\pi ^{\left( 0 \right)}}\left( {{{\left( {{f^{\left( 0 \right)}}} \right)}^2} + {{\left( {{\phi ^{\left( 0 \right)}}} \right)}^2}} \right)} \right\}} \right.
		\nonumber\\
		&\left. { + \frac{{\left( {1 - b{\rho _0}{\pi ^{\left( 0 \right)}}} \right){g^{\left( 1 \right)}} - b{\rho _0}{g^{\left( 0 \right)}}{\pi ^{\left( 1 \right)}}}}{{\left( {\gamma  - 1} \right)}}} \right]{x}dx,
	\end{align}
	\begin{align}\label{35}
		{\vartheta _2}{J_0} = \int_0^1&{\left[ {\frac{\gamma }{{2{{\left( {1 - b{\rho _0}} \right)}}}}\left\{ {2{\pi ^{\left( 0 \right)}}\left( {{f^{\left( 0 \right)}}{f^{\left( 2 \right)}} + {\phi ^{\left( 0 \right)}}{\phi ^{\left( 2 \right)}}} \right) + {\pi ^{\left( 0 \right)}}\left( {{{\left( {{f^{\left( 1 \right)}}} \right)}^2} + {{\left( {{\phi ^{\left( 1 \right)}}} \right)}^2}} \right)} \right\}} \right.}
		\nonumber\\
		&+ \frac{\gamma }{{2{{\left( {1 - b{\rho _0}} \right)}}}}\left\{ { 2{\pi ^{\left( 1 \right)}}\left( {{f^{\left( 0 \right)}}{f^{\left( 1 \right)}} + {\phi ^{\left( 0 \right)}}{\phi ^{\left( 1 \right)}}} \right) + {\pi ^{\left( 2 \right)}}\left( {{{\left( {{f^{\left( 0 \right)}}} \right)}^2} + {{\left( {{\phi ^{\left( 0 \right)}}} \right)}^2}} \right)} \right\}
		\nonumber\\
		&\left. { + \frac{{\left( {1 - b{\rho _0}{\pi ^{\left( 0 \right)}}} \right){g^{\left( 2 \right)}} + \left( {1 - b{\rho _0}{\pi ^{\left( 1 \right)}}} \right){g^{\left( 1 \right)}} + \left( {1 - b{\rho _0}{\pi ^{\left( 2 \right)}}} \right){g^{\left( 0 \right)}}}}{{\left( {\gamma  - 1} \right)}}} \right]{x}dx,
	\end{align}
	and so on.
	Now, using \eqref{33} in \eqref{27}, we obtain
	\begin{align}\label{36}
		y{\left( {\frac{{{r_{{s_0}}}}}{{{r_s}}}} \right)^{2}} =&{J_0}\left[ {1 - \frac{\gamma }{{4{J_0}{{\left( {1 - b{\rho _0}} \right)}}}}{{{\left( {\frac{{{v^*}}}{{{A^*}}}} \right)}^2}}}  + \left\{ {{\vartheta _1} - \frac{\left( {1 - b{\rho _0}} \right)}{{2{J_0}\left( {\gamma  - 1} \right)}}} \right\}y \right.
		\nonumber\\
		&\left. {+ {\vartheta _2}{y^2} + {\vartheta _3}{y^3} + ...} \right],
	\end{align}
	In view of Eq. \eqref{20}, Eq. \eqref{36} transforms into the following form:
	\begin{align}\label{37}
		{\left( {\frac{{{a_0}}}{U}} \right)^2}{\left( {\frac{{{r_{{s_0}}}}}{{{r_s}}}} \right)^{2}} =& {J_0}\left[ {1 - \frac{\gamma }{{4{J_0}{{\left( {1 - b{\rho _0}} \right)}}}} {{{\left( {\frac{{{v^*}}}{{{A^*}}}} \right)}^2}}}+ \left\{ {{\vartheta _1} - \frac{\left( {1 - b{\rho _0}} \right)}{{2{J_0}\left( {\gamma  - 1} \right)}}} \right\}{{\left( {\frac{{{a_0}}}{U}} \right)}^2} \right.
		\nonumber\\
		&\left. { + {\vartheta _2}{{\left( {\frac{{{a_0}}}{U}} \right)}^4} + {\vartheta _3}{{\left( {\frac{{{a_0}}}{U}} \right)}^6} + ...} \right].
	\end{align}
	The above equation in the form of power series in ($\left( {\frac{{{a_0}}}{U}} \right)^2$) provides the relationship between the velocity of the shock $U$ and the position of the shock front $r_{s}$. Since
	$y=\left( {\frac{{{a_0}}}{U}} \right)^2 \ll 1$ and the series expansions in Eqs. \eqref{31}, \eqref{32} and \eqref{36} are convergent. In order to verify this, we must find the higher order approximation solutions such as zeroth, first and so forth by using numerical techniques. Evaluating the first-, second-, or third-order solution using the numerical integration method is quite difficult. Hence, the goal of the current study is to find the analytical solution. We determined the analytical solution for the zeroth-order solution.''  Using Eq.\eqref{29} and \eqref{32}, the expression of $\lambda$ can be written as:
	\begin{equation}\label{38}
		\lambda  = 2\left[ {1 + \frac{{\left( {{\vartheta _1} - \frac{\left( {1 - b{\rho _0}} \right)}{{2{J_0}\left( {\gamma  - 1} \right)}}} \right)y}}{{1 - \frac{\gamma }{{4{J_0}{{\left( {1 - b{\rho _0}} \right)}}}}{{{\left( {\frac{{{v^*}}}{{{A^*}}}} \right)}^2}}}} + \frac{{2{\vartheta _2}{y^2}}}{{1 - \frac{\gamma }{{4{J_0}{{\left( {1 - b{\rho _0}} \right)}}}} {{{\left( {\frac{{{v^*}}}{{{A^*}}}} \right)}^2}}}} + ...} \right].
	\end{equation}
	Further using
	\begin{equation}\label{39}
		{\lambda _1} = \frac{{\left( {{\vartheta _1} - \frac{\left( {1 - b{\rho _0}} \right)}{{2{J_0}\left( {\gamma  - 1} \right)}}} \right)}}{{1 - \frac{\gamma }{{4{J_0}{{\left( {1 - b{\rho _0}} \right)}}}}{{{\left( {\frac{{{v^*}}}{{{A^*}}}} \right)}^2}}}},\quad
		{\lambda _2} = \frac{{2{\vartheta _2}}}{{1 - \frac{\gamma }{{4{J_0}{{\left( {1 - b{\rho _0}} \right)}}}} {{{\left( {\frac{{{v^*}}}{{{A^*}}}} \right)}^2}}}},
	\end{equation}
	and so on, Eq.\eqref{38} can be written as
	\begin{equation}\label{40}
		\lambda  = 2\left( {1 + {\lambda _1}y + {\lambda _2}{y^2} + ...} \right).
	\end{equation}
	To obtain the ODEs for the functions ``$f^{\left(k\right)}$, $\pi^{\left(k\right)}$, $g^{\left(k\right)}$ and $\phi^{\left(k\right)}$, $(k = 0, 1, 2,...)$, we use the Eqs. \eqref{31} and \eqref{40} in Eqs. \eqref{23}-\eqref{26} and compare the coefficients of like powers of $y$ on both sides. So, the comparisons of coefficients of zeroth power of $y$ on both sides give:
	\begin{align}
		&\left( {{f^{\left( 0 \right)}} - x} \right)\pi _x^{\left( 0 \right)} + {\pi ^{\left( 0 \right)}}\left( {f_x^{\left( 0 \right)} + \frac{{{f^{\left( 0 \right)}}}}{x}} \right) = 0,\label{41}\\
		&\left( {{f^{\left( 0 \right)}} - x} \right){\pi ^{\left( 0 \right)}}f_x^{\left( 0 \right)} + \frac{{\left( {1 - b{\rho _0}} \right)}}{\gamma }g_x^{\left( 0 \right)} - \left( {f^{\left( 0 \right)}} + \frac{{{{\left( {{\phi ^{\left( 0 \right)}}} \right)}^2}}}{x} \right){\pi ^{\left( 0 \right)}} = 0,\label{42}\\
		&\left( {{f^{\left( 0 \right)}} - x} \right)g_x^{\left( 0 \right)} + \frac{{\gamma {g^{\left( 0 \right)}}}}{{\left( {1 - b{\rho _0}{\pi ^{\left( 0 \right)}}} \right)}}\left( {f_x^{\left( 0 \right)} + \frac{{{f^{\left( 0 \right)}}}}{x}} \right) - 2{g^{\left( 0 \right)}} = 0,\label{43}\\
		&\left( {{f^{\left( 0 \right)}} - x} \right)\phi _x^{\left( 0 \right)} + \left( {\frac{{{f^{\left( 0 \right)}}}}{x} - 1} \right){\phi ^{\left( 0 \right)}} = 0.\label{44}
	\end{align}
	\\
	Now, the comparisons of coefficients of first power of $y$ on both sides give:
	\begin{align}
		&{\pi ^{\left( 0 \right)}}f_x^{\left( 1 \right)} + \left( {{f^{\left( 0 \right)}} - x} \right)\pi _x^{\left( 1 \right)} + \left( {\pi _x^{\left( 0 \right)} + \frac{{{\pi ^{\left( 0 \right)}}}}{x}} \right){f^{\left( 1 \right)}}+ \left( {f_x^{\left( 0 \right)} + 2 + \frac{{{f^{\left( 0 \right)}}}}{x}} \right){\pi ^{\left( 1 \right)}} = 0,\label{45}\\
		&\left( {{f^{\left( 0 \right)}} - x} \right){\pi ^{\left( 0 \right)}}f_x^{\left( 1 \right)} + \frac{{\left( {1 - b{\rho _0}} \right)}}{\gamma }g_x^{\left( 1 \right)} + \left( {f_x^{\left( 0 \right)} + 1} \right){\pi ^{\left( 0 \right)}}{f^{\left( 1 \right)}} - {\lambda _1}{\pi ^{\left( 0 \right)}}{f^{\left( 0 \right)}}\nonumber
		\\
		&-\left( {f^{\left( 0 \right)} - \left( {{f^{\left( 0 \right)}} - x} \right)f_x^{\left( 0 \right)}} \right){\pi ^{\left( 1 \right)}} - \frac{{{{\left( {{\phi ^{\left( 0 \right)}}} \right)}^2}{\pi ^{\left( 1 \right)}} + 2{\phi ^{\left( 0 \right)}}{\phi ^{\left( 1 \right)}}{\pi ^{\left( 0 \right)}}}}{x}=0,\label{46}\\
		&\frac{{\gamma {g^{\left( 0 \right)}}}}{{\left( {1 - b{\rho _0}{\pi ^{\left( 0 \right)}}} \right)}}f_x^{\left( 1 \right)} + \left( {g_x^{\left( 0 \right)} + \frac{{\gamma {g^{\left( 0 \right)}}}}{{\left( {1 - b{\rho _0}{\pi ^{\left( 0 \right)}}} \right)x}}} \right){f^{\left( 1 \right)}}+ \left( {f_x^{\left( 0 \right)} + \frac{{{f^{\left( 0 \right)}}}}{x}} \right)\frac{{\gamma {g^{\left( 1 \right)}}}}{{\left( {1 - b{\rho _0}{\pi ^{\left( 0 \right)}}} \right)}}\nonumber
		\\
		& + \left( {{f^{\left( 0 \right)}} - x} \right)g_x^{\left( 1 \right)} + \left( {\gamma \left( {b{\rho _0}{\pi ^{\left( 1 \right)}} + \frac{{\left( {f_x^{\left( 1 \right)} + \frac{{{f^{\left( 1 \right)}}}}{x}} \right)}}{{\left( {1 - b{\rho _0}{\pi ^{\left( 0 \right)}}} \right)}}} \right) - 2{\lambda _1}} \right){g^{\left( 0 \right)}} = 0,\label{47}\\
		&\left( {{f^{\left( 0 \right)}} - x} \right)\phi _x^{\left( 1 \right)} + {f^{\left( 1 \right)}}\phi _x^{\left( 0 \right)} + \left( {1 + \frac{{{f^{\left( 0 \right)}}}}{x}} \right){\phi ^{\left( 1 \right)}} + \left( {\frac{{{f^{\left( 1 \right)}}}}{x} - {\lambda _1}} \right){\phi ^{\left( 0 \right)}} = 0.\label{48}
	\end{align}\\
	By applying Eq.\eqref{31} in shock jump conditions \eqref{28} and comparing the like powers of $y$ on both sides, for the zeroth powers of $y$," the boundary conditions are as follows:
	\begin{align}\label{49}
		&{f^{\left( 0 \right)}}\left( 1 \right) = \frac{2}{{\gamma  + 1}}\left( {1 - b{\rho _0}} \right),\quad
		{\pi ^{\left( 0 \right)}}\left( 1 \right) = \frac{{\gamma  + 1}}{{\gamma  - 1}}\left( {1 - \frac{{2b{\rho _0}}}{{\gamma  - 1}}} \right),\nonumber\\
		&{g^{\left( 0 \right)}}\left( 1 \right) = \frac{{2\gamma }}{{\gamma  + 1}},\quad
		{\phi ^{\left( 0 \right)}}\left( 1 \right) = \frac{{{v^*}}}{{{A^*}}}.
	\end{align}
	And for the first powers of $y$, the boundary conditions are as follows:
	\begin{align}\label{50}
		&{f^{\left( 1 \right)}}\left( 1 \right) = \frac{{ - 2}}{{\gamma  + 1}}\left( {1 - b{\rho _0}} \right),\quad
		{\pi ^{\left( 1 \right)}}\left( 1 \right) = \frac{{ - 2\left( {\gamma  + 1} \right)}}{{{{\left( {\gamma  - 1} \right)}^2}}}\left( {1 - b{\rho _0}} \right),\nonumber\\
		&{g^{\left( 1 \right)}}\left( 1 \right) =  -  {\frac{{\gamma  - 1}}{{\gamma  + 1}}},\quad
		{\phi ^{\left( 1 \right)}}\left( 1 \right) = 0.
	\end{align}
	
	Now, we have to solve the system of ordinary differential equations \eqref{41}-\eqref{44} together with the boundary conditions \eqref{49} to get the value of $\pi^{\left(0\right)}$, $f^{\left(0\right)}$, $g^{\left(0\right)}$ and $\phi^{\left(0\right)}$. For the solutions of shock waves in the following forms, we get a zeroth-order approximation as
	
	\begin{align}\label{51}
		&{\rho _1} = {\rho _0}\pi^{\left(0\right)} \left( {x} \right),\quad
		{u_1} = Uf^{\left(0\right)} \left( {x} \right),\nonumber\\
		&{p_1} = {p_0}{\left( {\frac{U}{{{a_0}}}} \right)^2}g^{\left(0\right)} \left( {x} \right),\quad
		{v_1} = U\phi^{\left(0\right)} \left( {x} \right),
	\end{align}
	From equation \eqref{39}, we obtain:
	\begin{equation}\label{52}
		{\left( {\frac{{{a_0}}}{U}} \right)^2}{\left( {\frac{{{r_{{s_0}}}}}{{{r_s}}}} \right)^{2}} ={J_0}\left[ {1 - \frac{\gamma }{{4{J_0}{{\left( {1 - b{\rho _0}} \right)}}}} {{{\left( {\frac{{{v^*}}}{{{A^*}}}} \right)}^2}}}\right].
	\end{equation}
	The values of $\pi^{\left(1\right)}$, $f^{\left(1\right)}$, $g^{\left(1\right)}$ and $\phi^{\left(1\right)}$ are needed to obtain the first-order approximate solutions, which are derived from the system of equations \eqref{45}-\eqref{48}. The values of $\pi^{\left(0\right)}$, $f^{\left(0\right)}$, $g^{\left(0\right)}$ and $\phi^{\left(0\right)}$ can be taken from zeroth-order approximation. The solutions for the first-order approximations have the following forms:
	\begin{align}\label{53}
		&{\rho _1} = {\rho _0}\left(\pi^{\left(0\right)} + {\left( {\frac{a_0}{U}} \right)^2}\pi^{\left(1\right)} \right),\quad
		{u_1} = U\left(f^{\left(0\right)}+{\left( {\frac{a_0}{U}} \right)^2}f^{\left(1\right)}\right),\nonumber\\
		&{p_1} = {p_0}{\left( {\frac{U}{{{a_0}}}} \right)^2}\left(g^{\left(0\right)}+ {\left( {\frac{a_0}{U}} \right)^2}g^{\left(1\right)}\right),\quad
		{v_1} = U\left(\phi^{\left(0\right)} +{\left( {\frac{a_0}{U}} \right)^2}\phi^{\left(1\right)}\right).
	\end{align}
	\section{\textbf{The Zeroth-Order Approximation}}
	Eqs. \eqref{41}-\eqref{44} can be rewritten as follows:
	\begin{align}
		&f_x^{\left( 0 \right)} + \frac{{\left( {1 - b{\rho _0}} \right)}}{{\gamma \left( {{f^{\left( 0 \right)}} - x} \right){\pi ^{\left( 0 \right)}}}}g_x^{\left( 0 \right)} = \frac{{\frac{{{{\left( {{\phi ^{\left( 0 \right)}}} \right)}^2}}}{x} + {f^{\left( 0 \right)}}}}{{\left( {{f^{\left( 0 \right)}} - x} \right)}},\label{54}
		\\
		&\frac{{\pi _x^{\left( 0 \right)}}}{{{\pi ^{\left( 0 \right)}}}} = \frac{{f_x^{\left( 0 \right)} + \frac{{{f^{\left( 0 \right)}}}}{x}}}{{x - {f^{\left( 0 \right)}}}},\label{55}
		\\
		&\frac{{g_x^{\left( 0 \right)}}}{{{g^{\left( 0 \right)}}}} = \frac{{\frac{\gamma }{{\left( {1 - b{\rho _0}{\pi ^{\left( 0 \right)}}} \right)}}\left( {f_x^{\left( 0 \right)} + \frac{{{f^{\left( 0 \right)}}}}{x}} \right) - 2}}{{x - {f^{\left( 0 \right)}}}},\label{56}
		\\
		&\frac{{\phi _x^{\left( 0 \right)}}}{{{\phi ^{\left( 0 \right)}}}} = \frac{{\frac{{{f^{\left( 0 \right)}}}}{x} - 1}}{{x - {f^{\left( 0 \right)}}}}.\label{57}
	\end{align}
	Using the above equations, we obtained the expression for $f_x^{\left( 0\right)}$ as follows:
	
	\begin{equation}\label{58}
		f_x^{\left( 0 \right)} = \frac{{\left( {f^{\left( 0 \right)} + \frac{{{{\left( {{\phi ^{\left( 0 \right)}}} \right)}^2}}}{x}} \right)\frac{{\pi ^{\left( 0 \right)}}\left( {{f^{\left( 0 \right)}} - x} \right)}{g ^{\left( 0 \right)}} + \frac{{1 - b{\rho _0}}}{\gamma }\left( {\frac{{\gamma {f^{\left( 0 \right)}}}}{{\left( {1 - b{\rho _0}{\pi ^{\left( 0 \right)}}} \right)x}} - 2} \right)}}{{\frac{{{{\left( {{f^{\left( 0 \right)}} - x} \right)}^2}{\pi ^{\left( 0 \right)}}}}{{{g^{\left( 0 \right)}}}} - \frac{{1 - b{\rho _0}}}{{\left( {1 - b{\rho _0}{\pi ^{\left( 0 \right)}}} \right)}}}}.
	\end{equation}
	From Eqs.\eqref{49} and \eqref{58}, we have
	\begin{align}\label{59}
		f_x^{\left( 0 \right)}\left( 1 \right) = &\left[ {\frac{{\left( {1 - \gamma  - 2{b}{\rho _0}} \right)}}{{2\gamma \left( {\gamma  - 1} \right)}}\left( {1 - \frac{{2{b}{\rho _0}}}{{\gamma  - 1}}} \right)\left\{ {2\left( {1 - b{\rho _0}} \right) + {{\left( {\frac{{{v^*}}}{{{A^*}}}} \right)}^2}\left( {\gamma  + 1} \right)} \right\}} \right.\nonumber
		\\
		&{{\left. { + \frac{{\left( {1 - b{\rho _0}} \right)}}{\gamma }\left\{ {\frac{{2\gamma \left( {\gamma  - 1} \right)\left( {1 - b{\rho _0}} \right)}}{{\left( {\gamma  + 1} \right)\left\{ {\gamma  - 1 - b{\rho _0}\left( {\gamma  + 1} \right)\left( {1 - \frac{{2{b}{\rho _0}}}{{\gamma  - 1}}} \right)} \right\}}} - 2} \right\}} \right]} \mathord{\left/
				{\vphantom {{\left. { + \frac{{\left( {1 - \bar b} \right)}}{\gamma }\left\{ {\frac{{2\gamma \left( {\gamma  - 1} \right)\left( {1 - \bar b} \right)}}{{\left( {\gamma  + 1} \right)\left\{ {\gamma  - 1 - \bar b\left( {\gamma  + 1} \right)\left( {1 - \frac{{2\bar b}}{{\gamma  - 1}}} \right)} \right\}}} - \left( {m + 1} \right)} \right\}} \right]} }} \right.
				\kern-\nulldelimiterspace}}\nonumber
		\\
		&\left[ {\frac{{{{\left( {1 - \gamma  - 2{b}{\rho _0}} \right)}^2}}}{{2\gamma \left( {\gamma  - 1} \right)}}\left( {1 - \frac{{2{b}{\rho _0}}}{{\gamma  - 1}}} \right) - \frac{{\left( {\gamma  - 1} \right)\left( {1 - b{\rho _0}} \right)}}{{\left\{ {\gamma  - 1 - b{\rho _0}\left( {\gamma  + 1} \right)\left( {1 - \frac{{2{b}{\rho _0}}}{{\gamma  - 1}}} \right)} \right\}}}} \right].
	\end{align}
	From the similarity transformations, Zhuravskaya et al. [25] found the approximate solution to an intense
	explosion. The same approximation was used by Sakurai [6] to get analytical solutions in gas dynamics.
	Thus, we consider the following form of $f^{\left(0\right)}\left(x\right)$ by Taylor ( [3] and [4]):
	\begin{equation}\label{60}
		{f^{\left( 0 \right)}}\left( x \right) = \frac{x}{\gamma } + B{x^n},
	\end{equation}
	where $B$ and $n$ are constants which are determined by $f^{\left(0\right)}$ and $f_x^{\left( 0 \right)}$ given in the equations \eqref{49} and \eqref{59}. Thus, we obtain
	\begin{align}\label{61}
		B = \frac{{\gamma  - 1 - 2{b}{\rho _0}\gamma }}{{\gamma \left( {\gamma  + 1} \right)}},\quad
		n = \left( {f_x^{\left( 0 \right)}\left( 1 \right) - \frac{1}{\gamma }} \right)\frac{{\gamma \left( {\gamma  + 1} \right)}}{{\gamma  - 1 - 2{b}{\rho _0}\gamma }}.
	\end{align}
	Now by substituting the values of ${f^{\left( 0 \right)}}\left( x \right)$, $B$ and $n$ from equations \eqref{60} and \eqref{61}, respectively, into equations \eqref{55}-\eqref{57}, then integrating these equations and using the boundary conditions given in Eq.\eqref{49} to determine the integration constants, we obtained
	\begin{align}
		{\pi ^{\left( 0 \right)}}\left( x \right) =&\frac{{\gamma  + 1}}{{\gamma  - 1 + 2b{\rho _0}}}{x^{\frac{2}{{\gamma  - 1}}}}{\left( {\frac{{{\gamma ^2} - \gamma  + 2\gamma b{\rho _0}}}{{\gamma  - 1 - 2\gamma b{\rho _0}}}} \right)^{\frac{{2 - \left( {n + 1} \right)\left( {1 - \gamma } \right)}}{{\left( {1 - \gamma } \right)\left( {1 - n} \right)}}}}\times\nonumber
		\\
		&{\left( {\frac{{\left( {g - 1} \right)\left( {g + 1} \right)}}{{\gamma  - 1 - 2\gamma b{\rho _0}}} - {x^{n - 1}}} \right)^{\frac{{2 - \left( {n + 1} \right)\left( {1 - \gamma } \right)}}{{\left( {1 - \gamma } \right)\left( {n - 1} \right)}}}},\label{62}
		\\
		{g^{\left( 0 \right)}}\left( x \right) =&\frac{{2\gamma }}{{\gamma  + 1}}{\left( {1 - \frac{{b{\rho _0}\left( {\gamma  + 1} \right)}}{{\gamma  - 1 + 2b{\rho _0}}}} \right)^\gamma }\left( {1 - \frac{{b{\rho _0}\left( {\gamma  + 1} \right)}}{{\gamma  - 1 + 2b{\rho _0}}}{{{\left( {\frac{{{\gamma ^2} - \gamma  + 2\gamma b{\rho _0}}}{{\gamma  - 1 - 2\gamma b{\rho _0}}}} \right)}^{\frac{{2 - \left( {n + 1} \right)\left( {1 - \gamma } \right)}}{{\left( {1 - \gamma } \right)\left( {1 - n} \right)}}}}}\times }\right.\nonumber
		\\
		&{\left. {x^{\frac{2}{{\gamma  - 1}}}}{{{\left( {\frac{{\left( {g - 1} \right)\left( {g + 1} \right)}}{{\gamma  - 1 - 2\gamma b{\rho _0}}} - {x^{n - 1}}} \right)}^{\frac{{2 - \left( {n + 1} \right)\left( {1 - \gamma } \right)}}{{\left( {1 - \gamma } \right)\left( {n - 1} \right)}}}}} \right)^\gamma }{\left( {\frac{{{\gamma ^2} - \gamma  + 2\gamma b{\rho _0}}}{{\gamma  - 1 - 2\gamma b{\rho _0}}}} \right)^{\frac{{\gamma \left( {n + 1} \right)}}{{\left( {n - 1} \right)}}}}\times\nonumber
		\\
		&{\left( {\frac{{\left( {g - 1} \right)\left( {g + 1} \right)}}{{\gamma  - 1 - 2\gamma b{\rho _0}}} - {x^{n - 1}}} \right)^{\frac{{\gamma \left( {n + 1} \right)}}{{\left( {1 - n} \right)}}}}\label{63},
		\\
		{\phi ^{\left( 0 \right)}}\left( x \right) =&\frac{{{v^*}}}{{{A^*}}}x^{-1}.\label{64}
	\end{align}
	The zeroth-order approximate analytical solution is given by the equations \eqref{60} and \eqref{62}-\eqref{64} for the considered problem.
	\section{\textbf{The First-Order Approximation}}
	For the first-order approximations ${f}^{\left(1\right)}$, ${\pi}^{\left(1\right)}$, ${g}^{\left(1\right)}$ and ${\phi}^{\left(1\right)}$ of the flow variables $f$, $\pi$, $g$ and $\phi$, respectively, we shall derive the system of Ordinary Differential Equations. With the help of system of Differential Equations \eqref{45}-\eqref{48} and the boundary conditions \eqref{50}, we find these flow variables. Splitting the variables $f^{\left(1\right)}$, $\pi^{\left(1\right)}$, $g^{\left(1\right)}$ and $\phi^{\left(1\right)}$ as follows: 
	\begin{align}\label{65}
		&{f^{\left( 1 \right)}} = f_1^{\left( 1 \right)} + {\lambda _1}{f_2^{\left( 1 \right)}},\quad
		{\pi ^{\left( 1 \right)}} = \pi _1^{\left( 1 \right)} + {\lambda _1}\pi _2^{\left( 1 \right)},\nonumber\\
		&{g^{\left( 1 \right)}} = g_1^{\left( 1 \right)} + {\lambda _1}g_2^{\left( 1 \right)},\quad
		{\phi ^{\left( 1 \right)}} = \phi _1^{\left( 1 \right)} + {\lambda _1}\phi _2^{\left( 1 \right)},
	\end{align}	
	By substituting the values given in \eqref{65} into \eqref{45}-\eqref{48} we get two sets of equations for ${f_1}^{\left(1\right)}$, ${\pi_1}^{\left(1\right)}$, ${g_1}^{\left(1\right)}$, ${\phi_1}^{\left(1\right)}$ and ${f_2}^{\left(1\right)}$, ${\pi_2}^{\left(1\right)}$, ${g_2}^{\left(1\right)}$, ${\phi_2}^{\left(1\right)}$ which will be independent of $\lambda_1$.
	Now, comparisons of the zeroth-powers of $\lambda_1$ give the following set of equations:  
	\begin{align}
		&{\pi ^{\left( 0 \right)}}f_{1x}^{\left( 1 \right)} + \left( {{f^{\left( 0 \right)}} - x} \right)\pi _{1x}^{\left( 1 \right)} + \left( {\pi _x^{\left( 0 \right)} + \frac{{{\pi ^{\left( 0 \right)}}}}{x}} \right)f_1^{\left( 1 \right)} + \left({2 + f_x^{\left( 0 \right)} + \frac{f^{\left( 0 \right)}}{x}} \right)\pi _1^{\left( 1 \right)} = 0,\label{66}
		\\
		&\left( {{f^{\left( 0 \right)}} - x} \right)f_{1x}^{\left( 1 \right)}{\pi ^{\left( 0 \right)}} + \frac{{1 - b{\rho_{0}}}}{\gamma }g_{1x}^{\left( 1 \right)} + \left( {f_x^{\left( 0 \right)} + 1} \right)f_1^{\left( 1 \right)}{\pi ^{\left( 0 \right)}}\nonumber
		\\
		&- \left\{ {\left( {x - {f^{\left( 0 \right)}}} \right)f_x^{\left( 0 \right)} + {f^{\left( 0 \right)}}} \right\}\pi _1^{\left( 1 \right)} - \frac{{2{\pi ^{\left( 0 \right)}}{\phi ^{\left( 0 \right)}}\phi _1^{\left( 1 \right)} + {{\left( {{\phi ^{\left( 0 \right)}}} \right)}^2}\pi _1^{\left( 1 \right)}}}{x}=0,\label{67}
		\\
		&\left( {{f^{\left( 0 \right)}} - x} \right)g_{1x}^{\left( 1 \right)} + \frac{\gamma }{{{{\left( {1 - b{\rho_{0}}{\pi ^{\left( 0 \right)}}} \right)}^2}}}\left[ {b{\rho_{0}}\pi _1^{\left( 1 \right)}{g^{\left( 0 \right)}}\left( {f_x^{\left( 0 \right)} + \frac{f^{\left( 0 \right)}}{x}} \right)} \right.\nonumber
		\\
		&\left. { + \left( {1 - b{\rho_{0}}{\pi ^{\left( 0 \right)}}} \right)\left\{ {{g^{\left( 0 \right)}}\left( {f_{1x}^{\left( 1 \right)} + \frac{{f_1^{\left( 1 \right)}}}{x}} \right) + g_1^{\left( 1 \right)}\left( {f_x^{\left( 0 \right)} + \frac{{{f^{\left( 0 \right)}}}}{x}} \right)} \right\}} \right] =0,\label{68}
		\\
		&f_1^{\left( 1 \right)}\phi _x^{\left( 0 \right)} + \left( {{f^{\left( 0 \right)}} - x} \right)\phi _{1x}^{\left( 1 \right)} + \left( {1 + \frac{{{f^{\left( 0 \right)}}}}{x}} \right)\phi _1^{\left( 1 \right)} + \frac{{f_1^{\left( 1 \right)}{\phi ^{\left( 0 \right)}}}}{x} = 0.\label{69}
	\end{align}
	Again, comparisons of the first powers of $\lambda_1$ give the following set of equations:
	\begin{align}
		&{\pi ^{\left( 0 \right)}}f_{2x}^{\left( 1 \right)} + \left( {{f^{\left( 0 \right)}} - x} \right)\pi _{2x}^{\left( 1 \right)} + \left( {\pi _x^{\left( 0 \right)} + \frac{{{\pi ^{\left( 0 \right)}}}}{x}} \right)f_2^{\left( 1 \right)} + \left( {2+ f_x^{\left( 0 \right)} + \frac{f^{\left( 0 \right)}}{x}} \right)\pi _2^{\left( 1 \right)} = 0,\label{70}
		\\
		&\left( {{f^{\left( 0 \right)}} - x} \right)f_{2x}^{\left( 1 \right)}{\pi ^{\left( 0 \right)}} + \frac{{1 - b{\rho_{0}}}}{\gamma }g_{2x}^{\left( 1 \right)} + \left( {f_x^{\left( 0 \right)} + 1} \right)f_2^{\left( 1 \right)}{\pi ^{\left( 0 \right)}}- {f^{\left( 0 \right)}}{\pi ^{\left( 0 \right)}}\nonumber
		\\ 
		&- \left\{ {\left( {x - {f^{\left( 0 \right)}}} \right)f_x^{\left( 0 \right)} +f^{\left( 0 \right)}} \right\}\pi _2^{\left( 1 \right)} - \frac{{2{\pi ^{\left( 0 \right)}}{\phi ^{\left( 0 \right)}}\phi _2^{\left( 1 \right)} + {{\left( {{\phi ^{\left( 0 \right)}}} \right)}^2}\pi _2^{\left( 1 \right)}}}{x}=0,\label{71}
		\\
		&\left( {{f^{\left( 0 \right)}} - x} \right)g_{2x}^{\left( 1 \right)} + \frac{\gamma }{{{{\left( {1 - b{\rho_{0}}{\pi ^{\left( 0 \right)}}} \right)}^2}}}\left[ {b{\rho_{0}}\pi _2^{\left( 1 \right)}{g^{\left( 0 \right)}}\left( {f_x^{\left( 0 \right)} + \frac{f^{\left( 0 \right)}}{x}} \right) - 2{g^{\left( 0 \right)}}} \right.\nonumber
		\\
		&\left. { + \left( {1 - b{\rho_{0}}{\pi ^{\left( 0 \right)}}} \right)\left\{ {{g^{\left( 0 \right)}}\left( {f_{2x}^{\left( 1 \right)} + \frac{{f_2^{\left( 1 \right)}}}{x}} \right) + g_2^{\left( 1 \right)}\left( {f_x^{\left( 0 \right)} + \frac{f^{\left( 0 \right)}}{x}} \right)} \right\}} \right] = 0,\label{72}
		\\
		&f_2^{\left( 1 \right)}\phi _x^{\left( 0 \right)} + \left( {{f^{\left( 0 \right)}} - x} \right)\phi _{2x}^{\left( 1 \right)} + \left( {1 + \frac{{{f^{\left( 0 \right)}}}}{x}} \right)\phi _2^{\left( 1 \right)} + \frac{{f_2^{\left( 1 \right)}{\phi ^{\left( 0 \right)}}}}{x} - {\phi ^{\left( 0 \right)}} = 0.\label{73}
	\end{align}
	Now, substituting the values of variables given in \eqref{65} into the boundary conditions given in \eqref{50} and then by comparing the coefficients of zeroth and first powers of $\lambda_1$, we obtain 
	\begin{align}\label{74}
		&{{f_{1}}^{\left( 1 \right)}}\left( 1 \right) = \frac{{ - 2}}{{\gamma  + 1}}\left( {1 - b{\rho_{0}}} \right),\quad
		{{\pi_{1}} ^{\left( 1 \right)}}\left( 1 \right) = \frac{{ - 2\left( {\gamma  + 1} \right)}}{{{{\left( {\gamma  - 1} \right)}^2}}}\left( {1 - b{\rho_{0}}} \right),\nonumber\\
		&{{g_{1}}^{\left( 1 \right)}}\left( 1 \right) =  - \left( {\frac{{\gamma  - 1}}{{\gamma  + 1}}} \right),\quad
		{{\phi_{1}} ^{\left( 1 \right)}}\left( 1 \right) = 0,
	\end{align}
	and
	\begin{align}\label{75}
		&{{f_{2}}^{\left( 1 \right)}}\left( 1 \right) = 0,\quad
		{{\pi_{2}} ^{\left( 1 \right)}}\left( 1 \right) =0,\quad
		{{g_{2}}^{\left( 1 \right)}}\left( 1 \right) = 0,\quad
		{{\phi_{2}} ^{\left( 1 \right)}}\left( 1 \right) = 0,
	\end{align}
	\setlength{\arrayrulewidth}{.1mm}
	\setlength{\tabcolsep}{15pt}
	\renewcommand{\arraystretch}{1.5}
	\begin{table}[h]
		\caption{The values of $B$, $n$ and the zeroth approximation $J_0$ of $J$ for $\rho_0=1$ and varying values of the parameters $\gamma$, $b$, $\frac{{{v^*}}}{{{A^*}}}$.}
		\renewcommand{\arraystretch}{1.3}
		\label{table_1}
		\centering
		\scalebox{0.68}
		{\begin{tabular}{ccccccccc}
				\hline
				\multirow{3}{*}{$\gamma$}&\multirow{3}{*}{$b$}&\multirow{3}{*}{$B$}&\multicolumn{2}{c}{Non- Rotating Case}&\multicolumn{2}{c}{Rotating Case}&\multicolumn{2}{c}{Rotating Case}\\
				&&&\multicolumn{2}{c}{$\frac{{{v^*}}}{{{A^*}}}=0$}&\multicolumn{2}{c}{$\frac{{{v^*}}}{{{A^*}}}=0.5$}&\multicolumn{2}{c}{$\frac{{{v^*}}}{{{A^*}}}=1$}\\
				\cline{4-9}
				&&&$n$&$J_{0}$&$n$&$J_{0}$&$n$&$J_{0}$\\
				\hline
				\multirow{3}{*}{$1.33$}&$0$&$0.106489$&$7.91898$&$1.02023$\text{(Sakurai\cite{7})}&$10.2666$&$1.11725$&$17.3096$&$1.42385$\text{(Nath\cite{50})}\\
				&$0.0009$&$0.105717$&$7.83094$&$1.01665$&$10.1828$&$1.11389$&$17.2385$&$1.42108$\\
				&$0.0011$&$0.105545$&$7.81135$&$1.01586$&$10.1642$&$1.11315$&$17.2227$&$1.42047$\\
				\hline
				\multirow{3}{*}{$1.40$}&$0$&$0.119048$&$6.83333$&$0.878679$\text{(Sakurai\cite{7})}&$8.9333$&$0.98488$&$15.2333$&$1.32283$\text{(Nath\cite{50})}\\
				&$0.0009$&$0.118298$&$6.77026$&$0.875564$&$8.8740$&$0.98199$&$15.1855$&$1.32054$\\
				&$0.0011$&$0.118131$&$6.75623$&$0.874875$&$8.8608$&$0.98134$&$15.1748$&$1.32004$\\
				\hline
				\multirow{3}{*}{$1.667$}&$0$&$0.150026$&$4.74841$&$0.601236$\text{(Sakurai\cite{7})}&$6.41478$&$0.74781$&$11.4139$&$1.22367$\text{(Nath\cite{50})}\\
				&$0.0009$&$0.149351$&$4.7214$&$0.598991$&$6.39079$&$0.74582$&$11.3989$&$1.22242$\\
				&$0.0011$&$0.149201$&$4.71539$&$0.598494$&$6.38545$&$0.74538$&$11.3956$&$1.22215$\\
				\hline
		\end{tabular}}
	\end{table}
	\break
	With the help of boundary conditions \eqref{74}, \eqref{75} and the zeroth-order approximations of flow variables given in \eqref{60} and \eqref{62}-\eqref{64}, the system of Eqs. \eqref{66}-\eqref{69} gives the solution for ${f_1}^{\left(1\right)}$, ${\pi_1}^{\left(1\right)}$, ${g_1}^{\left(1\right)}$, ${\phi_1}^{\left(1\right)}$ and the system of Eqs. \eqref{70}-\eqref{73} gives the solution for ${f_2}^{\left(1\right)}$, ${\pi_2}^{\left(1\right)}$, ${g_2}^{\left(1\right)}$, ${\phi_2}^{\left(1\right)}$, respectively. 
	Now by substituting the values of ${f_1}^{\left(1\right)}$, ${\pi_1}^{\left(1\right)}$, ${g_1}^{\left(1\right)}$, ${\phi_1}^{\left(1\right)}$ and ${f_2}^{\left(1\right)}$, ${\pi_2}^{\left(1\right)}$, ${g_2}^{\left(1\right)}$, ${\phi_2}^{\left(1\right)}$ in Eq. \eqref{34}, we obtain the following expression for $\lambda_1$:
	\begin{equation}
		{\lambda _1} = \frac{{{I_1} - \frac{{1 - b{\rho_{0}}}}{{2\left( {\gamma  - 1} \right)}}}}{{\frac{J_0}{2} - \frac{\gamma }{{8\left( {1 -b{\rho_{0}}} \right)}} {{{\left( {\frac{{{v^*}}}{{{A^*}}}} \right)}^2}} - {I_2}}},\label{76}\\
	\end{equation}
	where
	\begin{align}
		{I_1} =&\int_0^1 {\left[ {\frac{\gamma }{{\left( {1 - b{\rho_{0}}} \right)}}\left\{ {{f^{\left( 0 \right)}}{\pi ^{\left( 0 \right)}}f_1^{\left( 1 \right)} + {\phi ^{\left( 0 \right)}}{\pi ^{\left( 0 \right)}}\phi _1^{\left( 1 \right)}} \right\}}+ \frac{{\gamma \pi _1^{\left( 1 \right)}}}{{2\left( {1 -b{\rho_{0}}} \right)}}\left\{ {{{\left( {{f^{\left( 0 \right)}}} \right)}^2} + {{\left( {{\phi ^{\left( 0 \right)}}} \right)}^2}} \right\} \right.}\nonumber
		\\
		&\left.{+ \frac{1}{{\gamma  - 1}}\left\{ {\left( {1 - b{\rho_{0}}{\pi ^{\left( 0 \right)}}} \right)g_1^{\left( 1 \right)} - b{\rho_{0}}\pi _1^{\left( 1 \right)}{g^{\left( 0 \right)}}} \right\}} \right]{x}dx,\label{77}
	\end{align}
	and
	\begin{align}
		{I_2} =&\int_0^1 { \left[ {\frac{\gamma }{{\left( {1 - b{\rho_{0}}} \right)}}\left\{ {{f^{\left( 0 \right)}}{\pi ^{\left( 0 \right)}}f_2^{\left( 1 \right)} + {\phi ^{\left( 0 \right)}}{\pi ^{\left( 0 \right)}}\phi _2^{\left( 1 \right)}} \right\}}+ \frac{{\gamma \pi _2^{\left( 1 \right)}}}{{2\left( {1 - b{\rho_{0}}} \right)}}\left\{ {{{\left( {{f^{\left( 0 \right)}}} \right)}^2} + {{\left( {{\phi ^{\left( 0 \right)}}} \right)}^2}} \right\}\right.}\nonumber
		\\
		&\left.{ + \frac{1}{{\gamma  - 1}}\left\{ {\left( {1 - b{\rho_{0}}{\pi ^{\left( 0 \right)}}} \right)g_2^{\left( 1 \right)} - b{\rho_{0}}\pi _2^{\left( 1 \right)}{g^{\left( 0 \right)}}} \right\}} \right]{x}dx.\label{78}
	\end{align}
	Using the above values of ${f_1}^{\left(1\right)}$, ${\pi_1}^{\left(1\right)}$, ${g_1}^{\left(1\right)}$, ${\phi_1}^{\left(1\right)}$, ${f_2}^{\left(1\right)}$, ${\pi_2}^{\left(1\right)}$, ${g_2}^{\left(1\right)}$, ${\phi_2}^{\left(1\right)}$ and $\lambda_1$, the values of first-order approximation ${f}^{\left(1\right)}$, ${\pi}^{\left(1\right)}$, ${g}^{\left(1\right)}$ and ${\phi}^{\left(1\right)}$ to the flow variables $f, \pi, g, \phi$ can be calculated by Eqs. \eqref{65}.
	\section{\textbf{Results and Discussion}}
	
\begin{figure*}[htp]
		\centering
		\subfigure[Flow pattern of Density]{\includegraphics[scale=0.50]{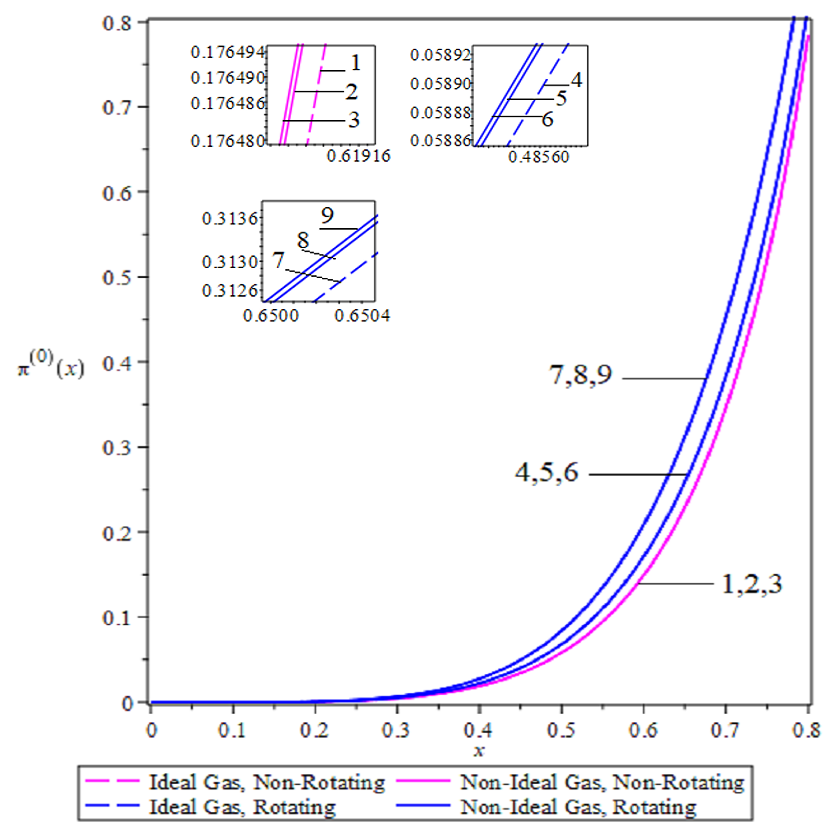}}\quad\label{fig1a}
		\subfigure[Flow pattern of Radial Fluid Velocity]{\includegraphics[scale=0.50]{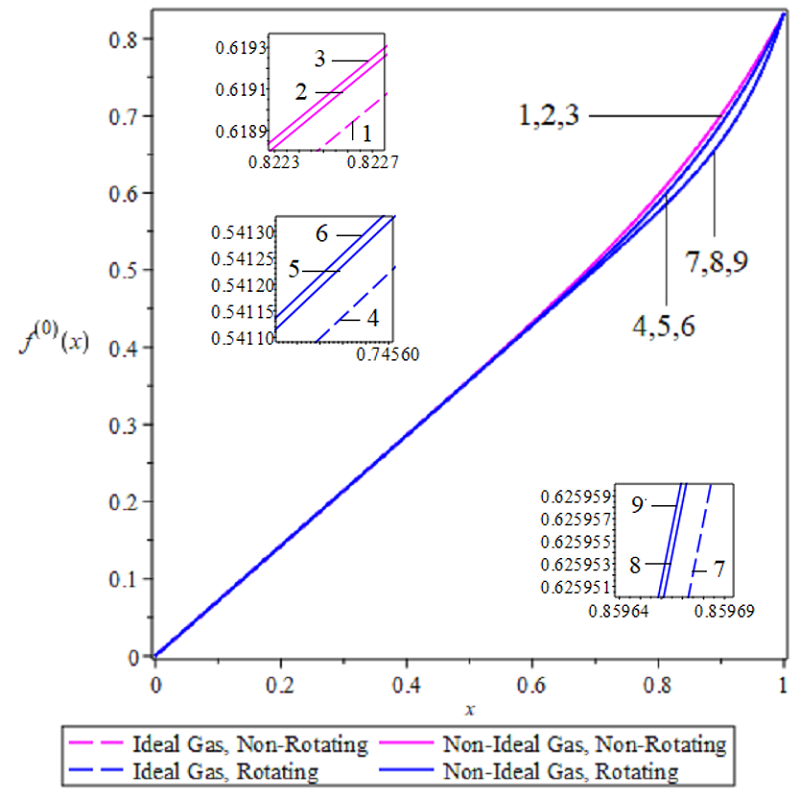}}\label{fig1b}\\
		\subfigure[Flow pattern of Pressure]{\includegraphics[scale=0.50]{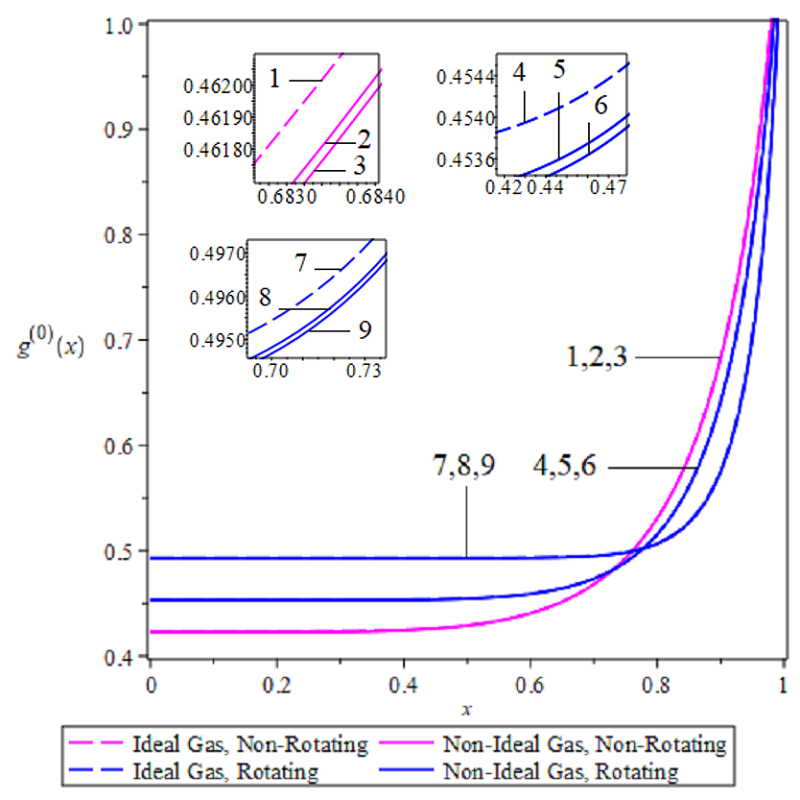}}\quad\label{fig1c}
		\subfigure[Flow pattern of Azimuthal Velocity]{\includegraphics[scale=0.50]{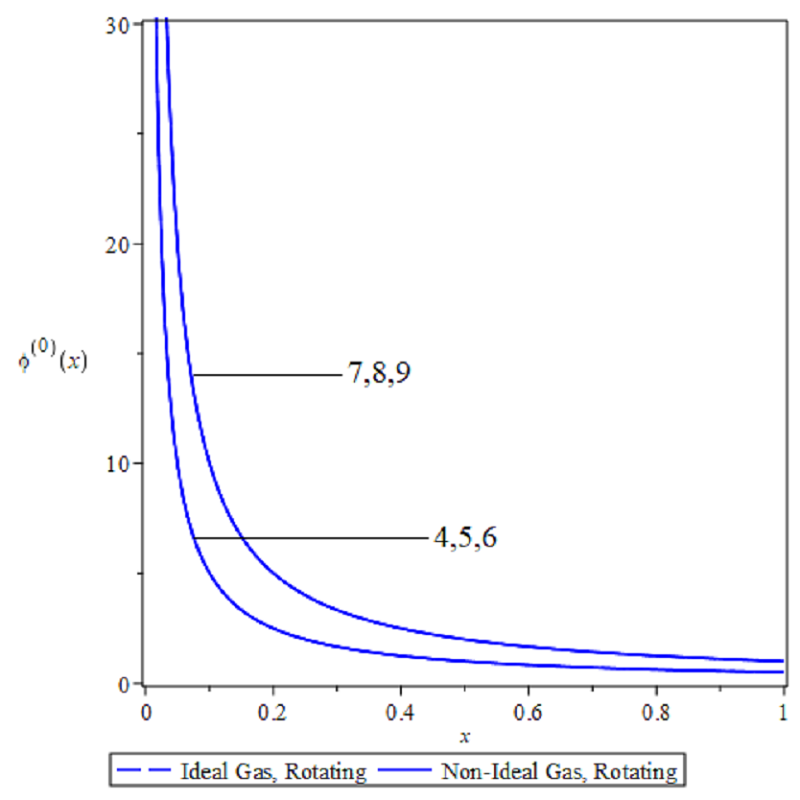}}\label{fig1d}\\
		\caption{ Flow patterns behind the cylindrical shock front of reduced flow variables for ${\frac{{{v^*}}}{{{A^*}}}=0}$ (non-rotating case ), ${\frac{v^*}{A^*}=0.5,1}$ (rotating case) and $\gamma={1.40}$ with different values of $b=0$ (Ideal Gas, i.e., dash curves) and $b=0.0009,0.0011$ (Non-Ideal Gas, i.e., solid curves): 1. $\gamma={1.40}$, $b=0$, $\frac{v^*}{A^*}=0$; 2. $\gamma={1.40}$, $b=0.0009$, $\frac{v^*}{A^*}=0$; 3. $\gamma={1.40}$, $b=0.0011$, $\frac{v^*}{A^*}=0$; 4. $\gamma={1.40}$, $b=0$, $\frac{v^*}{A^*}=0.5$; 5. $\gamma={1.40}$, $b=0.0009$, $\frac{v^*}{A^*}=0.5$; 6. $\gamma={1.40}$, $b=0.0011$, $\frac{v^*}{A^*}=0.5$; 7. $\gamma={1.40}$, $b=0$, $\frac{v^*}{A^*}=1$; 8. $\gamma={1.40}$, $b=0.0009$, $\frac{v^*}{A^*}=1$; 9. $\gamma={1.40}$, $b=0.0011$, $\frac{v^*}{A^*}=1$.}
		\label{fig1}
	\end{figure*}

\begin{figure*}[htp]
		\centering
		\subfigure[Flow pattern of Density]{\includegraphics[scale=0.50]{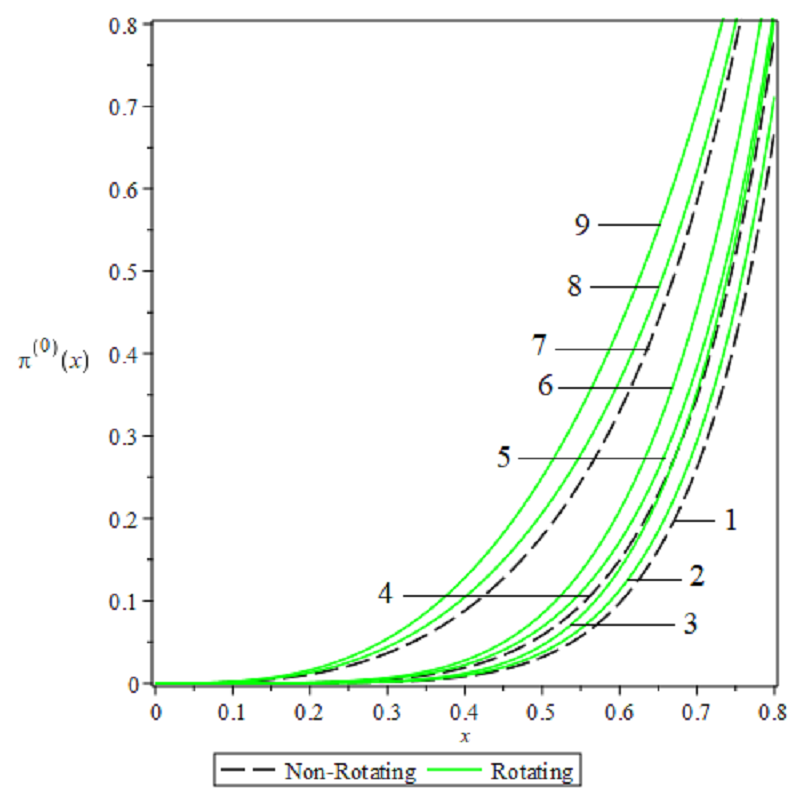}}\quad\label{fig2a}
		\subfigure[Flow pattern of Radial Fluid Velocity]{\includegraphics[scale=0.50]{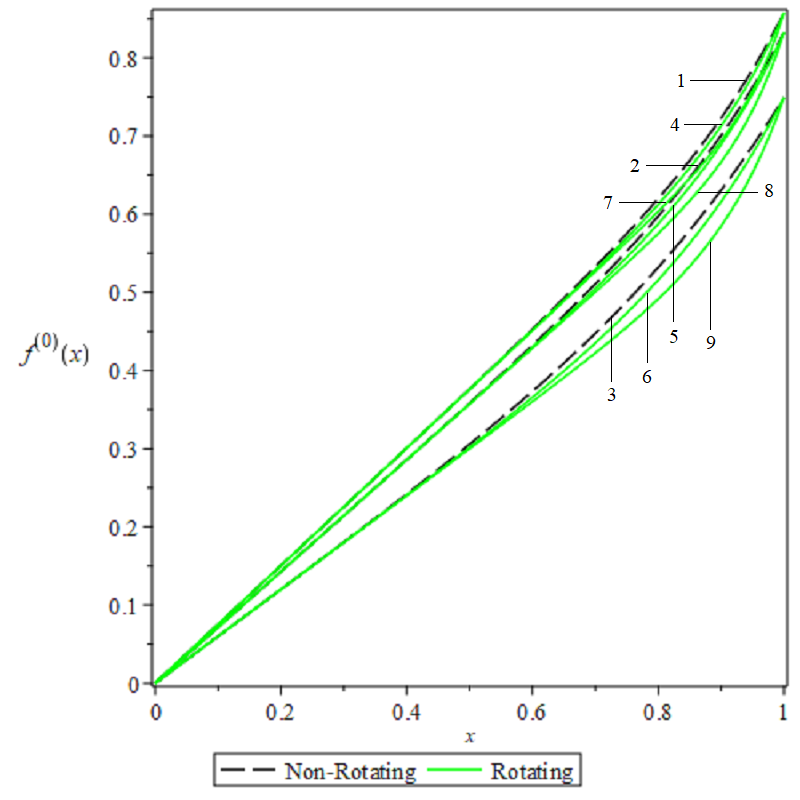}}\label{fig2b}\\
		\subfigure[Flow pattern of Pressure]{\includegraphics[scale=0.50]{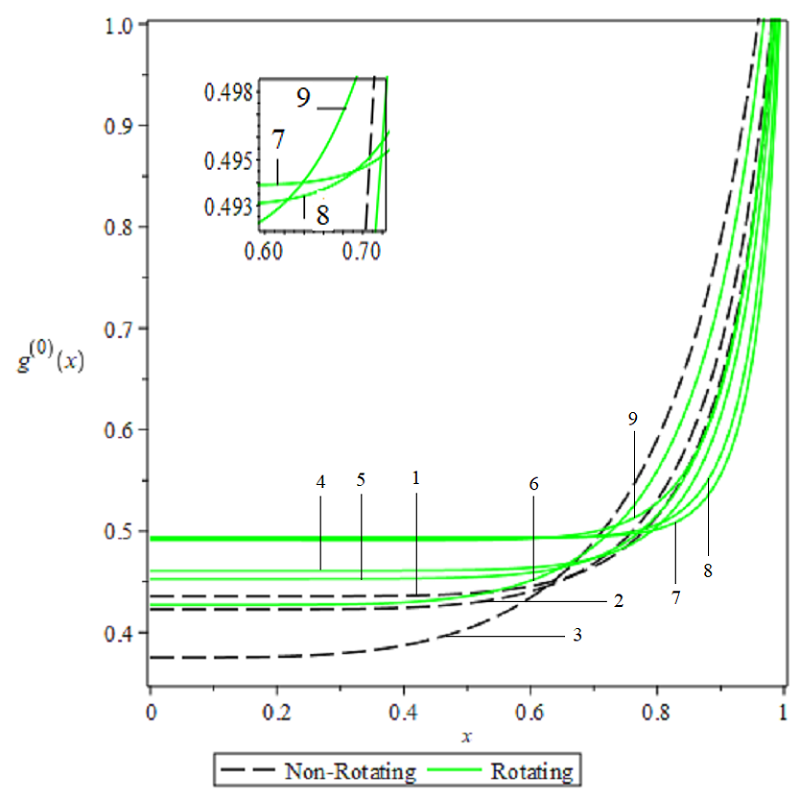}}\quad\label{fig2c}
		\subfigure[Flow pattern of Azimuthal Velocity]{\includegraphics[scale=0.50]{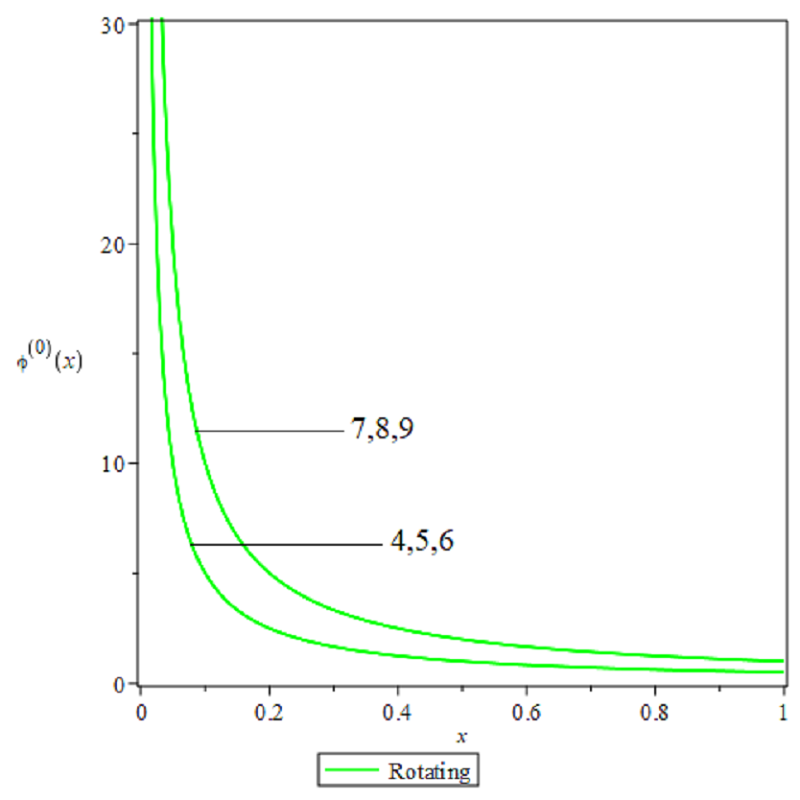}}\label{fig2d}\\
		\caption{ Flow patterns behind the cylindrical shock front of reduced flow variables for ${\frac{{{v^*}}}{{{A^*}}}=0}$ (non-rotating case, i.e., dash curves), ${\frac{v^*}{A^*}=0.5,1}$ (rotating case, i.e., solid curves) and non-ideal gas parameter $b=0.0011$  with different values of $\gamma=1.33, 1.40, 1.667$: 1. $\gamma={1.33}$, $b=0.0009$, $\frac{v^*}{A^*}=0$; 2. $\gamma={1.40}$, $b=0.0009$, $\frac{v^*}{A^*}=0$; 3. $\gamma={1.667}$, $b=0.0009$, $\frac{v^*}{A^*}=0$; 4. $\gamma={1.33}$, $b=0.0009$, $\frac{v^*}{A^*}=0.5$; 5. $\gamma={1.40}$, $b=0.0009$, $\frac{v^*}{A^*}=0.5$; 6. $\gamma={1.667}$, $b=0.0009$, $\frac{v^*}{A^*}=0.5$; 7. $\gamma={1.33}$, $b=0.0009$, $\frac{v^*}{A^*}=1$; 8. $\gamma={1.40}$, $b=0.0009$, $\frac{v^*}{A^*}=1$; 9. $\gamma={1.667}$, $b=0.0009$, $\frac{v^*}{A^*}=1$.}
		\label{fig2}
	\end{figure*}

	The current studies presents the effect on the shock propagation and flow variables of the non-idealness of the gas in a rotating medium. For the case of zeroth-order approximation, the distributions of the flow variables density , radial fluid velocity, pressure and azimuthal fluid velocity denoted by $\pi ^{\left( 0 \right)}\left( x \right)$, $f ^{\left( 0 \right)}\left( x \right)$, $g ^{\left( 0 \right)}\left( x \right)$ and $\phi ^{\left( 0 \right)}\left( x \right)$ respectively, are drawn using the approximate analytical solution given in \eqref{60} and \eqref{62}-\eqref{64} for the cylindrical symmetric flows, and are shown in Figs. \ref{fig1} and \ref{fig2}. The constants $B$, $n$, and the zeroth-order approximation of $J_0$ in rotating and non-rotating medium are displayed in Table \ref{table_1}.The numerical values of the approximate analytic solutions given in Eqs. \eqref{60}, \eqref{62} and \eqref{63} are depicted in Table \ref{table_2}. The values of the physical parameters assumed for numerical computation operations are
	\\
	
	$\gamma=1.33, 1.40, 1.667$ \cite{50}; $b=0, 0.0009, 0.0011$ \cite{37}; $\frac{{{v^*}}}{{{A^*}}}=0, 0.5, 1$ \cite{50}; $\rho_0=1$.
	\\
	\setlength{\arrayrulewidth}{.1mm}
	\setlength{\tabcolsep}{15pt}
	\renewcommand{\arraystretch}{1.5}
	\begin{table}[h]
		\caption{Values of the flow variables $f^{\left( 0 \right)}$, $\pi^{\left( 0 \right)}$ and $g^{\left( 0 \right)}$ for $\gamma=1.40$, $\frac{{{v^*}}}{{{A^*}}}=0$ and $\rho_0=1$.}
		\renewcommand{\arraystretch}{1.5}
		\label{table_2}
		\footnotesize\setlength{\tabcolsep}{15pt}
		\centering
		\scalebox{0.8}
		{\begin{tabular}{|p{0.5cm}|p{0.8cm}|p{1cm}|p{1cm}|p{2cm}|}
				\hline
				$x$&$b$&$f^{\left( 0 \right)}$&$\pi^{\left( 0 \right)}$&$g^{\left( 0 \right)}$\\		
				\hline
				\multirow{3}{*}{1}&0&0.8333&6.000&1.667\text{(Sakurai\cite{7})}\\
				&0.0009&0.8325&5.973&1.149\\
				&0.0011&0.8324&5.967&1.145\\
				\multirow{3}{*}{0.9}&0&0.7008&1.898&0.685\text{(Sakurai\cite{7})}\\
				&0.0009&0.7008&1.901&0.682\\
				&0.0011&0.7008&1.902&0.682\\
				\multirow{3}{*}{0.8}&0&0.5973&0.783&0.531\text{(Sakurai\cite{7})}\\
				&0.0009&0.5975&0.784&0.530\\
				&0.0011&0.5975&0.784&0.530\\
				\multirow{3}{*}{0.7}&0&0.5104&0.347&0.468\text{(Sakurai\cite{7})}\\
				&0.0009&0.5105&0.346&0.468\\
				&0.0011&0.5106&0.346&0.468\\
				\multirow{3}{*}{0.6}&0&0.4322&0.149&0.441\text{(Sakurai\cite{7})}\\
				&0.0009&0.4322&0.149&0.440\\
				&0.0011&0.4323&0.149&0.440\\
				\multirow{3}{*}{0.5}&0&0.3582&0.058&0.429\text{(Sakurai\cite{7})}\\
				&0.0009&0.3582&0.058&0.428\\
				&0.0011&0.3582&0.058&0.428\\
				\multirow{3}{*}{0.4}&0&0.2859&0.019&0.425\text{(Sakurai\cite{7})}\\
				&0.0009&0.2859&0.018&0.424\\
				&0.0011&0.2859&0.018&0.424\\
				\multirow{3}{*}{0.3}&0&0.2143&0.005&0.424\text{(Sakurai\cite{7})}\\
				&0.0009&0.2143&0.004&0.423\\
				&0.0011&0.2143&0.004&0.422\\
				\multirow{3}{*}{0.2}&0&0.1429&0.001&0.424\text{(Sakurai\cite{7})}\\
				&0.0009&0.1428&0.000&0.422\\
				&0.0011&0.1428&0.000&0.422\\
				\multirow{3}{*}{0.1}&0&0.0714&0.000&0.424\text{(Sakurai\cite{7})}\\
				&0.0009&0.0714&0.000&0.422\\
				&0.0011&0.0714&0.000&0.422\\
				\hline
		\end{tabular}}
	\end{table}
	\break	
	The effect of blast (shock) wave  propagation for cylindrical symmetry on the flow variables due to varying the values of non-ideal parameter $b$, adiabatic exponent $\gamma$ and the rotational parameters $\frac{{{v^*}}}{{{A^*}}}$, i.e., for rotating and non-rotating cases are shown in Figures \ref{fig1} and \ref{fig2}, respectively. In the present work, we have extended the works of Sakurai \cite{7} and Nath \cite{50} by considering the rotating medium for non-ideal gas while Sakurai has considered non-rotating medium for ideal gas and Nath's solution is for rotating ideal gas. The values $b=0$ and $\frac{{{v^*}}}{{{A^*}}}=0$ correspond to the ideal and non-rotating case, respectively (i.e., the solution given by Sakurai \cite{7}) which is depicted in the curve 1 of Figs. \ref{fig1}(a-c). Curve 4 and curve 7 of Figs. \ref{fig1}(a-d) are for ideal rotating gas which shows the result given by Nath \cite{50}, remaining curves are new of Fig. \ref{fig1} in which the non-zero values are taken for the non-ideal parameter $b$. And Fig. \ref{fig2} shows the effect of adiabatic exponent on the flow parameter. The Numerical values for different values of $b$ with $\gamma=1.4$ and $\frac{{{v^*}}}{{{A^*}}}=0$ for cylindrical symmetry of the functions $\pi ^{\left( 0 \right)}$, $f ^{\left( 0 \right)}$, $g ^{\left( 0 \right)}$ and $\phi ^{\left( 0 \right)}$  are presented in Table \ref{table_2}. From Table \ref{table_2}, it is quite evident that the obtained results recover Sakurai's results \cite{7} very well for the case of an ideal gas. Figs. \ref{fig1} (a-c), shows that the density $\pi ^{\left( 0 \right)}\left( x \right)$, radial fluid velocity $f ^{\left( 0 \right)}\left( x \right)$ and pressure $g ^{\left( 0 \right)}\left( x \right)$ increase as we move towards the shock front from axis of symmetry. These flow variables have maximum value at the shock front and minimum values near the axis of symmetry. From Fig.\ref{fig1} (d), it is shown that the azimuthal fluid velocity $\phi ^{\left( 0 \right)}\left( x \right)$ increases as we move towards the axis of symmetry from the shock front, and it has minimum value at the shock front and maximum value near the axis of symmetry. The values of $J_0$ from the Eq. \eqref{33} are calculated numerically using the values of the zeroth-order approximation to the solution of flow variables $\pi ^{\left( 0 \right)}\left( x \right)$, $f ^{\left( 0 \right)}\left( x \right)$, $g ^{\left( 0 \right)}\left( x \right)$,  $\phi ^{\left( 0 \right)}\left( x \right)$ are shown in the Table \ref{table_1}. Also, the calculated values of $J_{0}$ are compared with the computed values given by Sakurai\cite{7} and Nath\cite{50} for non-rotating and rotating ideal gas, respectively, and it seems that the values of $J_{0}$ are in close agreement with  Sakurai\cite{7} and Nath\cite{50}. From Eq. \eqref{52}, it is clear that the shock velocity decreases as $J_0$ increases, i.e., the shock strength decreases and vice versa.\\ 
	\\
	\textbf{7.1 Effects of Non-Ideal parameter ${b}$}\\
	
	The effect of the non-ideal parameter $b$ on the flow variables when adiabatic exponent $\gamma$ is kept fixed for both rotating and non-rotating medium can be seen in Fig. \ref{fig1}. Also, the effects of $b$ on the constant quantities are shown in Table \ref{table_1}. We observed that with an increase in the value of non-ideal parameter $b$, density  $\pi ^{\left( 0 \right)}\left( x \right)$ and radial fluid velocity $f ^{\left( 0 \right)}\left( x \right)$ increase behind the shock (see Fig. \ref{fig1}(a-b)), whereas from Fig. \ref{fig1}(c) and  Table \ref{table_1}, it is observed that pressure $g ^{\left( 0 \right)}\left( x \right)$ and the constant quantities $B$, $n$ and $J_0$ decrease. Also, it is shown in the Fig. \ref{fig1}(d) that the azimuthal fluid velocity is unaffected by increasing the value of $b$.\\
	\break
	
	\textbf{7.2 Effects of Adiabatic exponent $\gamma$}\\
	
	The effect of the adiabatic exponent $\gamma$ on the flow variables when the parameter of non-ideal gas $b$ is kept fixed for both the non-rotating and rotating motion can be seen in Fig. \ref{fig2}. Also, the effects of $\gamma$ on the constant quantities are shown in Table \ref{table_1}. We observed that with an increase in the value of adiabatic exponent $\gamma$, the flow variable radial fluid velocity $f ^{\left( 0 \right)}\left( x \right)$ and the constant quantities $n$ and $J_0$ decrease while $B$ increases (see Fig. \ref{fig2}(b) and Table \ref{table_1}). From Fig.\ref{fig2}(a,c), it is observed that the density  $\pi ^{\left( 0 \right)}\left( x \right)$ increases near the axis of symmetry but it decreases while moving towards the shock front whereas pressure $g ^{\left( 0 \right)}\left( x \right)$ shows the reverse effect. Also, the azimuthal components of fluid velocity $\phi ^{\left( 0 \right)}\left( x \right)$ remains unaffected as the value of $\gamma$ increases (see Fig. \ref{fig2}(d)). Also, the result obtained from Fig. \ref{fig2} are quite similar to the result obtained by Nath \cite{50} for perfect gas.\\
With increase in the value of adiabatic exponent $\gamma$, pressure increases near the shock front but as we move towards the axis of symmetry it gets decreased. Physically, it depicts that behind the shock front the gas is less compressed for a higher value of $\gamma$.
	\\
	\\ 
	\textbf{7.3 Effects of Rotational parameter $\frac{{{v^*}}}{{{A^*}}}$}\\
	
	Due to rotation, the density $\pi ^{\left( 0 \right)}\left( x \right)$ and pressure $g ^{\left( 0 \right)}\left( x \right)$ decrease near the shock front and increase towards the axis of symmetry and , azimuthal fluid velocity $\phi ^{\left( 0 \right)}\left( x \right)$, $J_0$ and $n$ increase; whereas the radial fluid velocity $f ^{\left( 0 \right)}\left( x \right)$ decreases (see Fig. \ref{fig2}(a-d) and Table \ref{table_1}). Also, the result obtained from Fig. \ref{fig2} are quite similar to the result obtained by Nath \cite{50} for perfect gas.\\
	The pressure decreases near the shock front and gets increased as we move towards the axis of symmetry. Physically, it depicts that due to the rotational motion, the gas is highly compressible behind the shock front for higher values of rotational parameter.
	
	\section{\textbf{Conclusions}}
	In this paper, we investigated ``the explosion problem for blast wave propagation in the rotating gas atmosphere for a $4\times4$ quasi-linear hyperbolic system of one-dimensional unsteady and inviscid cylindrically symmetric motion in van der Waals gas. Using the Sakurai's technique, the approximate analytical solutions for the zeroth-order of the flow variables such as density, radial fluid velocity, pressure, azimuthal fluid velocity are discussed in detail. Also, a brief description is given for obtaining the first-order approximation. Graphical representation is presented to describe the consequences of varying non-ideal parameters $b$, adiabatic exponent $\gamma$ and rotational parameter $\frac{{{v^*}}}{{{A^*}}}$ on the flow variables (see Fig. \ref{fig1} and \ref{fig2}). Table \ref{table_1} contains the numerical values for the constant quantities $B$, $n$, and the zeroth-approximation $J_0$ of $J$ for different values of the parameters like non-ideal parameters $b$, adiabatic exponent $\gamma$ and rotational parameter $\frac{{{v^*}}}{{{A^*}}}$ which are compared with the results obtained by Sakuari \cite{7} for ideal non-rotating gas ($b=0$ and $\frac{{{v^*}}}{{{A^*}}}=0$) and Nath \cite{50} for ideal rotating gas ($b=0$ and $\frac{{{v^*}}}{{{A^*}}}\neq 0$). Also, a comparison between the results of the present work for the non-ideal gas and the results of Sakurai \cite{7} for ideal gas for the flow variables density $\pi ^{\left( 0 \right)}\left( x \right)$, radial fluid velocity $f ^{\left( 0 \right)}\left( x \right)$ and pressure $g ^{\left( 0 \right)}\left( x \right)$ is listed in Table \ref{table_2}". In the absence of van der Waals excluded volume ($b=0$), our results match well with the Sakurai's and Nath's results. The distributions of the flow variables for the rotating non-ideal gas environment and the non-rotating ideal gas atmosphere show significant differences. \\
	The study of explosions is useful in the ionospheres, supernova explosions and blast waves generated due to strong explosions. The luminous trail of enormous length that follows the meteoroid is associated with strong cylindrical shock waves. Aside from the intense occurrences in active galactic nuclei, supernova explosions cause the strongest shock waves in galaxies. Important information about the energetics of the supernova event and the characteristics of the interstellar medium, where these shocks propagate, can be found in analyses of supernova remnants. The fundamental astrophysical process of star creation involves the interaction of blast waves with interstellar clouds \cite{48}.\\
	Thermal radiation has a substantial impact on the wave phenomena because the shock propagation processes are connected to a high-temperature gas dynamics phenomenon, due to its interaction with the magnetic field and its numerous theoretical and practical applications, an  analytical investigation  of  the  current  study for the mixture of non-ideal with the small solid particles and thermal radiation effects can be a potential future scope of the present work. Also, in the present work, we have only shown the Zeroth-order approximations but in future we can find higher-order approximations of the present model. So now, we can conclude as follows, considering the present
	work,
	\begin{enumerate}
		\item The figures (Fig.\ref{fig1} and Fig.\ref{fig2}) show that as we start from the shock front and approach the axis of symmetry, the flow variables like density, radial fluid velocity, and pressure drop, while the azimuthal component of radial fluid velocity decreases as we move from the axis of symmetry towards the shock front.
		\item From Figs. \ref{fig1}(d) and \ref{fig2}(d), we observe that the azimuthal fluid velocity is unaffected by varying the values of non-ideal gas parameter and adibatic exponent but it increases for rotating motion.
		\item With a rise in $b$, the flow density and velocity show increments, while pressure behaves in the opposite way. Physically, this result is expected because the charged gas particles are gradually carried away from the shock front. As a result, the gas particles are able to travel more freely, which increases the velocity of the flow behind the shock. (see Fig. \ref{fig1}(a, b, c)).   
		\item With constant $b$ and varying adiabatic exponent $\gamma$, we observed the behaviour of flow variables as shown in the Fig. \ref{fig2}(a,b,c). By increasing the value of $\gamma$, the radial fluid velocity decreases. Moving close to the axis of symmetry, the density increases, and it decreases while we move towards the shock front however the pressure behaves just opposite, it increases near the shock front and decreases while we approach the axis of symmetry.
		\item From Figs. \ref{fig1} and \ref{fig2}, we observed that close to the shock front, the flow density and pressure decrease, but they increase as we approach the axis of symmetry, and the azimuthal radial fluid velocity increases whereas radial fluid velocity behaves reversely due to rotation.
	\end{enumerate}
	\section{Acknowledgement}
	The author, Nandita, is grateful for the financial support provided by ``the Ministry of Education", New Delhi, India under the scheme of Senior Research Fellowship.
	
	\section{Conflict of Interest}
	There is no conflict of interest in this work.
	
	\section{Data Availability}
	The data that support the findings of this study are available
	within the article.
	
	\bibliography{Reference}
	\bibliographystyle{unsrt}
	
\end{document}